\documentclass[12pt]{article}
\newlength{\mytopmargin}
\newlength{\myleftmargin}
\newcommand{\mmu}{{\mathcal{M}}}
\newcommand{\J}{{\mathcal{J}}}
\newcommand{\I}{{\mathcal{I}}}
\newcommand{\erfc}{{\mathrm{erfc}}}
\newcommand{\OO}{\mathrm{O}}
\newcommand{\soft}{{\text{soft}}}
\newcommand{\sphere}{{\text{sphere}}}
\renewcommand{\r}{{\vec{r}}}
\newcommand{\ap}{{\text{approx}}}

\makeatletter
\@addtoreset{equation}{section}
\@addtoreset{table}{section}
\makeatother

\renewcommand{\theequation}{\thesection.\arabic{equation}}

\usepackage{amsmath,amsfonts,amssymb}

\usepackage{graphicx}

\begin{document}

\title{Expanded Vandermonde powers and sum rules for the two-dimensional one-component plasma}
\author{Gabriel T\'ellez${}^*$ and Peter J. Forrester${}^\dagger$}
\date{}
\maketitle

\noindent
\thanks{\small 
${}^*$Departamento de F\'{\i}sica, Universidad de Los Andes,
  Bogot\'a, Colombia\\
${}^\dagger$Department of Mathematics and Statistics, 
The University of Melbourne,
Victoria 3010, Australia
}

\begin{abstract}
\noindent The two-dimensional one-component plasma (2dOCP) is a system
of $N$ mobile particles of the same charge $q$ on a surface with a
neutralising background. The Boltzmann factor of the 2dOCP at temperature
$T$ can be expressed as a Vandermonde determinant to the power
$\Gamma=q^{2}/(k_B T)$. Recent advances in the theory of symmetric and
anti-symmetric Jack polymonials provide an efficient way to expand
this power of the Vandermonde in their monomial basis, allowing the
computation of several thermodynamic and structural properties of the
2dOCP for $N$ values up to 14 and   
$\Gamma$ equal to 4, 6 and 8. In this work, we explore two applications of this formalism to
study the moments of the pair correlation function of the 2dOCP on a
sphere, and the distribution of radial linear statistics of the 2dOCP
in the plane.
\end{abstract}

\noindent Key words: Coulomb gas; one-component plasma; Jack
polynomials; sum rules.

\section{Introduction}
Generally the two-dimensional one-component plasma (2dOCP) refers to a
model in classical statistical mechanics consisting of $N$ mobile
point particles of the same charge $q$ and a smeared out neutralising
background, confined in a two-dimensional surface. All the charges (point and continuous) interact via the solution $v(\vec{r},\vec{r}\,')$ of the Poisson equation on that surface. For the plane this is
\begin{equation}
\label{eq:Coulomb-plane}
v(\vec{r},\vec{r}\,') = - \log \Big ( |\vec{r} - \vec{r}\,'|/L \Big ),  
\end{equation}
where $L$ is an arbitrary length scale, which will be set equal to one
from now on, while for a sphere of radius $R$,
\begin{equation}\label{eq:Coulomb-sphere}
v((\theta,\phi),(\theta',\phi')) = - \log (2R |u'v - u v'|),
\end{equation}
where with $(\theta,\phi)$ and $(\theta',\phi')$ spherical coordinates,
$u,v$ (and similarly $u'$, $v'$) are the Cayley-Klein parameters
\begin{equation}\label{1.2}
u := \cos (\theta/2) e^{i \phi/2}, \qquad  v :=  - i \sin (\theta/2) e^{-i \phi/2}
\end{equation}
(see e.g.~\cite[\S 15.6.1]{Fo10}). In the latter case, due to the
rotational invariance of the sphere, the background-particle
interaction only contributes a constant, and the Boltzmann factor, at
temperature $T$, is then computed to be
\begin{equation}\label{15.110}
\Big ( {1 \over 2R} \Big )^{N \Gamma/2} e^{\Gamma N^2/4}
\prod_{1 \le j < k \le N} |u_k v_j - u_j v_k |^\Gamma,
\end{equation}
where $\Gamma = q^2/(k_B T)$. In the former case, the particles couple to the background via a harmonic
potential towards the origin, and the Boltzmann factor is then (see e.g.~\cite[eq.~(1.72)]{Fo10})
\begin{equation}\label{15.110a}
A_\Gamma 
e^{- \pi \Gamma \rho_b \sum_{j=1}^N |\vec{r}_j |^2/2}
\prod_{1 \le j < k \le N} |\vec{r}_k - \vec{r}_j  |^\Gamma, \qquad A_\Gamma =
e^{-\Gamma N^2 ((1/2) \log R - 3/8)},
\end{equation}
where $\rho_b$ is the density of the background. Although the derivation of (\ref{15.110a}) requires that the mobile particles and background
be confined to  a disk of
radius $R$, for our purposes below we will relax this constraint by
allowing the particles to locate anywhere in the plane. This situation
will be called the soft disk geometry.

Both (\ref{15.110a}) and (\ref{15.110}) have significance in a field of theoretical physics quite distinct
from that of classical plasmas. Consider for definiteness (\ref{15.110a}). According to the well known 
Vandermonde determinant identity (see e.g.~\cite[eq.~(1.173)]{Fo10}), for $\Gamma = 2$ and after introducing polar
coordinates $\vec{r} = (r,\theta)$ and then complex coordinates $z_j = r_j e^{i \theta_j}$, this can be written in the Slater determinant form
\begin{equation}\label{1.1}
 A_2 \Big | \det [ \phi_{j-1}(z_k) ]_{j,k=1,\dots,N}] \Big |^2, \qquad  \phi_{j-1}(z) =  z^{j-1} e^{- \pi 
\rho_b r^2}
\end{equation}
Moreover, the single particle wave functions $ \phi_{j-1}(z)$, $j=1,\dots,N$ are, with $\pi \rho_b = 1/4 l^2$
$l = \sqrt{\hbar c/e B}$ the magnetic length, ground state (lowest Landau level)
eigenfunctions of the Schr\"odinger equation for
particles confined to the plane in a perpendicular magnetic field of
strength $B$ (see e.g.~\cite[\S 15.2.1]{Fo10}). The integer $j$ has
the interpretation of being proportional  to the centre of the corresponding
classical cyclotron orbits. Thus (\ref{1.1}) is the absolute value squared of $N$ non-interacting fermions 
in the plane, subject to a perpendicular magnetic field which in turn is required to be sufficiently strong
that the spin degree of freedom of the fermions is frozen out. For 
$\Gamma/2$ an odd integer, it is a celebrated result of Laughlin \cite{La83}, that the (un-normalized) wave function
$$
 \Big ( \det [ \phi_{j-1}(z_k) ]_{j,k=1,\dots,N}] \Big )^{\Gamma/2} =
e^{- \pi  \rho_b  \Gamma \sum_{j=1}^N |\vec{r}_j |^2/2}
\prod_{1 \le j < k \le N} (z_k - z_j  )^{\Gamma/2}
$$
is an accurate trial wave function for the fractional quantum Hall effect in the case of filling fraction
$2/\Gamma$. Furthermore, the same holds for
$$
\prod_{1 \le j < k \le N} (u_k v_j - u_j v_k )^{\Gamma/2}
$$
in relation to fractional quantum Hall states on the sphere \cite{RR99}.

We have highlighted the plasma system in the plane, and on a sphere.
As is usual in statistical mechanics, bulk properties such as the free energy per particle are expected be independent of the boundary
conditions. In two-dimensions, the sphere is distinguished by the usual correction to the bulk free energy
per particle --- the surface tension --- being absent due to the homogeneity of the geometry. This
feature of the sphere makes it well suited for studies in which the aim is to extrapolate bulk values from
finite $N$ data \cite{CLWH82,HR85,TF99}. Extrapolation of finite $N$ data for the 2dOCP is to be a
primary concern of
this present paper, and as such we too will make use of spherical geometry and consider the 
plasma system specified by the Boltzmann factor (\ref{15.110}).

Specifically, we aim to probe the finite $N$ analogues of the known sum rules for the bulk two-particle
correlation function $\rho_{(2)}(\vec{r},\vec{0})$, or more conveniently the truncated bulk two-particle
correlation function $\rho_{(2)}^T(\vec{r},\vec{0}):= \rho_{(2)}(\vec{r},\vec{0}) - \rho^2$, where $\rho$
denotes the particle density. These sum rules read \cite{Ma88,KMST99}
\begin{align}
&{1 \over \rho} \int_{\mathbb R^2} \rho_{(2)}^T(\vec{r},\vec{0}) \, d \vec{r} = - 1 \label{1a}\\
& \int_{\mathbb R^2} r^2 \rho_{(2)}^T(\vec{r},\vec{0}) \, d \vec{r} = - {2 \over \pi \Gamma}  \label{1b}\\
& \rho  \int_{\mathbb R^2} r^4 \rho_{(2)}^T(\vec{r},\vec{0}) \, d \vec{r} = - {16 \over (\pi \Gamma)^2} \Big ( 1 - {\Gamma \over 4} \Big )  \label{1c}\\
&\rho^2  \int_{\mathbb R^2} r^6 \rho_{(2)}^T(\vec{r},\vec{0}) \, d \vec{r} = - {18 \over (\pi \Gamma)^3} \Big ( \Gamma -  6 \Big ) \Big ( \Gamma - {8 \over 3}  \Big ).   \label{1d}
\end{align}
It is straightforward to show from the definitions
that (\ref{1a}) is exact for all $N$ on the sphere. In fact the sphere plays no specific role; replacing
$\rho$ on the LHS by $\rho(\vec{0})$ this holds for a general one-component system.
Less obvious and special to the sphere geometry is that (\ref{1b}), appropriately modified, can
also be made exact for finite $N$. The derivation is given
in Section \ref{S3} below.

To probe (\ref{1c}) and (\ref{1d}) for finite $N$ we have available an exact expression for 
$ \rho_{(2)}^T(\vec{r},\vec{r}\,')$ at the special coupling $\Gamma = 2$ \cite{Ca81}. 
More generally, for $\Gamma$ an even integer  (\ref{15.110}) is the absolute value squared of
a multivariable polynomial. In the case that $\Gamma/2$ is also even,
a recent advance \cite{BH08a} identifies the polynomial in terms of the symmetric Jack polynomials (see
\cite{KK09} and \cite{Fo10} for textbook  treatments), while for $\Gamma/2$ odd an even more
recent paper \cite{BR09} shows that the polynomial is an anti-symmetric Jack polynomial. These Jack
polynomials have certain structural properties (related to their monomial basis expansion) which makes
them the most efficient known way to carry out exact numerical computations at these couplings.
We review the formalism of such computation methods in Section \ref{S2}, and we give too
an alternative derivation to the result of \cite{BR09}, \cite{TERB11}  for the eigenoperator of anti-symmetric Jack polynomials.
The results of calculations based on this formalism are reported in Section \ref{S3}. In Section \ref{S4}, an approximation to the moments is probed using a formalism based on the direct correlation function. 

One viewpoint on the moments of $\rho_{(2)}^T(\vec{r},\vec{0})$ is
that they are the averaged value of the linear statistic $\sum_{l=1}^N
|\vec{r}_l|^{2n}$ in the system perturbed by a particle being fixed at
the origin, minus the average value of this same linear statistic
before the perturbation. This suggests the question of computing
fluctuation formulas for linear statistics, and we take up this task
in Section \ref{S5}. Section \ref{S5} is independent from Sections
\ref{S3} and \ref{S4}. Some concluding remarks are made in Section
\ref{S6}.

\section{Preliminary material}\label{S2}
In this section the expansion formulas underlying the exact calculations of the moments --- both
analytic and numeric --- will be presented. Before doing so, for the sphere geometry, it is convenient to
map the particles to the plane by applying a stereographic projection. With the latter carried out by mapping
from the south pole to the plane tangent to the north pole, and with $(\theta, \phi)$ the usual spherical coordinates, this is specified by the equation
$$
z = 2 R e^{i \phi} \tan {\theta \over 2}, \qquad z = x + i y,
$$
and we then have
\begin{eqnarray}\label{QZ}
\lefteqn{ \Big ( {1 \over 2 R} \Big )^{ N \Gamma /2} e^{\Gamma N^2/2}
\prod_{1 \le j < k \le N} | u_k v_j - u_j v_k|^\Gamma d S_1 \cdots d S_N} \nonumber \\
&&  = 
 \Big ( {1 \over 2 R} \Big )^{ N \Gamma /2} e^{\Gamma N^2/2}
\prod_{j=1}^N {1 \over (1 + |z_j|^2/(4 R^2) )^{2 + \Gamma(N-1)/2}}
\prod_{1 \le j < k \le N}
\Big | {z_j - z_k \over 2 R} \Big |^\Gamma d \vec{r}_1 \cdots d \vec{r}_N.
\end{eqnarray}

As previously remarked, for $\Gamma$ even the product of differences in both (\ref{15.110a}) and
(\ref{15.110}) is the absolute value squared of a polynomial. This remains true of the stereographic projection
of (\ref{15.110}) as seen in (\ref{QZ}). In the case $\Gamma = 4p$ the polynomial is symmetric, while in the
case $\Gamma = 4p+2$ the polynomial is anti-symmetric. As a consequence, the two case need to be treated
separately.

\subsection{The case $\Gamma = 4p$}
Following \cite{TF99}, let $\mu = (\mu_1,\dots,\mu_N)$ be a partition of $pN (N-1)$ such that
\begin{equation}\label{2.1}
2 p (N - 1) \ge \mu_1 \ge \mu_2 \ge \cdots \ge \mu_N \ge 0,
\end{equation}
and, with $m_i$ denoting the corresponding frequency of the integer $i$ in the partition, define the 
corresponding monomial symmetric function by
$$
m_\mu(z_1,\dots,z_N) = {1 \over \prod_i m_i !}
\sum_{\sigma \in S_N} z_{\sigma(1)}^{\mu_1} \cdots z_{\sigma(N)}^{\mu_N}.
$$
We expand
\begin{equation}\label{2.1a}
\prod_{1 \le j < k \le N}
(z_k - z_j)^{2p} =
\sum_\mu c_\mu^{(N)}(2p) m_\mu(z_1,\dots,z_n).
\end{equation}
The significance of knowledge of $\{c_\mu^{(N)}(2p)\}$ in this expansion is that a simple calculation
shows \cite[eq.~(2.7)]{TF99}
\begin{eqnarray}\label{3.1}
I_{N,\Gamma}[g] & := & \int_{\mathbb R^2} d \vec{r}_1 \, g(r_1^2) \cdots
 \int_{\mathbb R^2} d \vec{r}_N \, g(r_N^2) \, \prod_{1 \le j < k \le N} |\vec{r}_k - \vec{r}_j|^\Gamma
\nonumber \\
& = & N! \pi^N \sum_\mu { (c_\mu^{(N)}(2p))^2 \over  \prod_i m_i !} \prod_{l=1}^N G_{\mu_l},
\end{eqnarray}
where
\begin{equation}\label{gg}
G_{\mu_l}[g] := 2 \int_0^\infty r^{1 + 2 \mu_l} g(r^2) \, dr.
\end{equation}

\subsection{The case $\Gamma = 4p+2$}
We again follow  \cite{TF99}. We now take $\mu$ to be a partition of $(p+1)N(N-1)$ such that
\begin{equation}\label{2.13a}
(2 p +1) (N - 1) \ge \mu_1 >  \mu_2 > \cdots > \mu_N \ge 0
\end{equation}
(note that the strict inequalities between the parts of the partition implies $m_i = 1$, in the notation used
below (\ref{2.1})). We then expand
\begin{equation}\label{3.1aa}
\prod_{1 \le j < k \le N} (z_k - z_j)^{2p +1} =
\sum_\mu c_\mu^{(N)}(2p+1) {\mathcal A} (z_1^{\mu_1} \cdots z_N^{\mu_N}),
\end{equation}
where $\mathcal A$ denotes anti-symmetrization. From the definition of the Schur polynomials
$s_\mu(z_1,\dots,z_n)$, and with $\delta_N := (N-1,N-2,\dots,0)$, this is equivalent to the expansion
\begin{equation}\label{3.1a}
\prod_{1 \le j < k \le N} (z_k - z_j)^{2p} =
\sum_\mu c_\mu^{(N)}(2p+1)  s_{\mu - \delta_N} (z_1,\dots,z_N),
\end{equation}
and now for the integral (\ref{3.1}) we have
\begin{equation}\label{3.1x}
I_{N,\Gamma}[g] = N! \pi^N \sum_\mu (c_\mu^{(N)}(2p+1))^2 \prod_{l=1}^N G_{\mu_l}.
\end{equation}

\subsection{The coefficients $c_\mu^{(N)}(2p)$ and $c_\mu^{(N)}(2p+1)$}
It has recently been observed \cite{BH08a,BR09} that the products in (\ref{2.1a}) and (\ref{3.1a}) can
be expressed in terms of the Jack symmetric polynomial $P_\kappa^{(\alpha)}(z)$ and Jack anti-symmetric
polynomial $S_\kappa^{(\alpha)}(z)$ respectively, where $z := (z_1,\dots,z_N)$.
Generally the nonsymmetric Jack polynomials are eigenfunctions of the differential operator
\cite[eq.~(11.63)]{Fo10}
\begin{eqnarray}\label{4}
\tilde{H}^{(C,Ex)} & = & \sum_{j=1}^N \Big ( z_j {\partial \over \partial z_j} \Big )^2 +
{N - 1 \over \alpha} \sum_{j=1}^N z_j {\partial \over \partial z_j} \nonumber \\
&& + {2 \over \alpha}
\sum_{1 \le j < k \le N}
{z_j z_k \over z_j - z_k}
\Big ( \Big ( {\partial \over \partial  z_j} -  {\partial \over \partial  z_k} \Big ) -
{ 1 - M_{jk} \over z_j - z_k} \Big )
\end{eqnarray}
where $M_{ij}$ interchanges $z_i$ and $z_j$. The operators of symmetrization and anti-symmetrization commute with this operator, and so with $M_{jk} = 1$ and $M_{jk} = -1$ respectively, (\ref{4}) is the
differential eigenoperator for $P_\kappa^{(\alpha)}(z)$ and $S_\kappa^{(\alpha)}(z)$.
This characterisation specifies the polynomials uniquely when supplemented by the structural formulas
\begin{eqnarray}
P_\kappa^{(\alpha)}(z) & = & m_\kappa(z) + \sum_{\rho < \kappa} c_{\kappa \rho}(\alpha)  m_\kappa(z)
\label{5.a} \\
S_\kappa^{(\alpha)}(z) & = & s_{\kappa - \delta_N}(z) +
 \sum_{\rho - \delta_N < \kappa - \delta_N } \tilde{c}_{\kappa \rho}(\alpha)  s_{\rho - \delta_N} (z),
\label{5.b} 
\end{eqnarray}
where in (\ref{5.b}) all parts of $\kappa$ are required to be distinct, and $\mu < \kappa$ refers to the
dominance partial ordering on partitions $|\mu| < |\kappa|$, specified by
$\sum_{i=1}^k \mu_i \le  \sum_{i=1}^k \kappa_i$, $(k=1,\dots,N)$.

We remark that in the work \cite{BR09} the anti-symmetric polynomials, defined as in \cite{BF97b} as the anti-symmetric Jack polynomials,
where not considered directly. Rather attention was focussed on $\prod_{1 \le i < j \le N} (z_i - z_j) P_\kappa^{(\alpha)}(z)$, which was
shown to be the polynomial eigenfunction of the differential operator
$$
\sum_{j=1}^N \Big ( z_j {\partial \over \partial z_j} \Big )^2 + {1 \over 2} \Big ( {1 \over \alpha} - 1 \Big )
\sum_{i \ne j} \Big ( {z_i + z_j \over z_i - z_j} \Big ( z_i  {\partial \over \partial z_j} -   z_j  {\partial \over \partial z_i}  \Big ) - 2
{z_i^2 + z_j^2 \over (z_i - z_j)^2} \Big ).
$$
Simple manipulation reduces this to
\begin{eqnarray}\label{5.1}
&& 
\sum_{j=1}^N \Big ( z_j {\partial \over \partial z_j} \Big )^2 +  \Big ( {1 \over \alpha} - 1 \Big )
\sum_{i \ne j} \Big ( {z_i z_j \over z_i - z_j} \Big (   {\partial \over \partial z_i} -    {\partial \over \partial z_j}  \Big ) 
- 2 {z_i z_j \over ( z_i - z_j)^2} \Big ) \nonumber \\
&& \qquad +  \Big ( {1 \over \alpha} - 1 \Big ) (N-1) \sum_{j=1}^N z_j  {\partial \over \partial z_j} -
 \Big ( {1 \over \alpha} - 1 \Big ) N (N - 1).
 \end{eqnarray}
 Comparison with (\ref{4}) reveals that up to a constant this is equal to $\tilde{H}^{(C,Ex)}$ with $\alpha$ replaced by
 $\alpha/(1 - \alpha)$ and $M_{ij} = - 1$, telling us that (\ref{5.1}) is an eigenoperator for $S_\kappa^{(\alpha/(1 - \alpha)}(z)$.
 This conclusion is consistent with the known relation \cite{BF97b}
 $$
S_\kappa^{(\alpha/(1 - \alpha))}(z) = \prod_{1 \le i < j \le N} (z_i - z_j) P_\kappa^{(\alpha)}(z).
$$

The products in (\ref{2.1a}) and (\ref{3.1a}) relate to the Jack polynomials with a negative parameter.
Explicitly we have \cite{BH08a} (see \cite{BF11} for a derivation based on earlier literature)
 \begin{equation}
 \prod_{1 \le j < k \le N} (z_k - z_j)^{2p} = P_{2p \delta_N}(z; -2/(2p-1))
 \end{equation}
 and \cite{BR09}
 \begin{equation}
 \prod_{1 \le j < k \le N} (z_k - z_j)^{2p+1} = P_{(2p+1) \delta_N}(z; -2/(2p+1)),
 \end{equation}
 where generally for $c \in \mathbb Z^+$, $c \kappa := (c \kappa_1,c \kappa_2,\dots, c \kappa_N)$. The computational
 advantage of these formulas in computing the coefficients $c_\mu^{(N)}$ (\ref{2.1a}) and (\ref{3.1a}) is that the
 partitions $\mu$ for which $c_\mu^{(N)}$ are nonzero is greatly restricted by the dominance ordering exhibited in
 (\ref{5.a}) and (\ref{5.b}). Furthermore the eigenfunction characterization allows recursive formulas for
 the coefficients in these formulas, and thus the coefficients $\{c_\mu^{(N)} \}$  to be obtained.
 
 In relation to (\ref{5.a}), let $\rho = (\rho_1,\rho_2,\dots,\rho_N)$, and let $\mu$ be constructed from $\rho$ by first
 adding $r$ to $\rho_i$ and subtracting $r$ from $\rho_j$ with the requirement that upon suitable reordering 
 to be a partition it
 satisfies $\rho < \mu \le \kappa$. Also define
 $$
 e_\kappa(\alpha) := \sum_{i=1}^N \kappa_i (\kappa_i - 1 - {2 \over \alpha} (i-1) ).
 $$
 Then \cite[\S VI.4 Example 3(d)]{Ma95}
 \begin{equation}
   \label{eq:cmu-boson}
 c_{\kappa \rho} = {1 \over e_\kappa(\alpha) - e_\rho(\alpha) }
 {2 \over \alpha} \sum_{\rho < \mu \le \kappa} ((\rho_i + r) - (\rho_j - r)) c_{\kappa \mu}(\alpha).
 \end{equation}

Regarding now (\ref{5.b}), let $\rho$ and $\mu$ be as above and define
$$
 e_\kappa^{\rm F}(\alpha)  :=  \sum_{i=1}^N \kappa_i (\kappa_i  + 2 i (  1 - {1\over \alpha} ) ).
 $$
 Further define $(-1)^{N_{\rm SW}}$ as the sign of the permutation needed to be applied to $\mu$ so that it is a partition.
 Then we have \cite{BR09, TERB11}
 \begin{equation}
   \label{eq:cmu-fermion}
 \tilde{c}_{\kappa \rho} = {1 \over e_\kappa^{\rm F}(\alpha) - e_\rho^{\rm F}(\alpha) }
 {2 \over \alpha} \sum_{\rho < \mu \le \kappa} (\rho_i -  \rho_j ) \tilde{c}_{\kappa \mu}(\alpha) (-1)^{N_{\rm SW}} .
 \end{equation}

\section{Moments of the correlation function on the sphere}\label{S3}

\subsection{Second moment}

The partition function of the 2dOCP on a sphere is~\cite{Ca81, TF99}
\cite[\S 15.6]{Fo10}
\begin{eqnarray}
Q&=&\frac{1}{N!}\int e^{-\beta U} dS_1 \ldots dS_N \\
&=&\frac{e^{\Gamma N^2/4}}{N! (2R)^{N^2\Gamma/2}}
  \int_{\mathbb{R}^{2N}} \prod_{1\leq j<k\leq N} |z_k-z_j|^{\Gamma} 
  \prod_{j=1}^{N} \frac{d^2\vec{r}_j
  }{\left(1+\frac{|z_j|^2}{(2R)^2}\right)^{2+(N-1)\frac{\Gamma}{2}}}
\,,
\end{eqnarray}
where $U$ is the potential energy of the 2dOCP and $z_j=r_{j}e^{i\phi_j}$ are the
coordinates of the stereographic projection of a sphere of radius
$R$. Equivalently, one can factor out the radius, introducing
$\tilde{z}_j=z_j/(2R)$, to obtain
\begin{equation}
Q
=\frac{e^{\Gamma N^2/4}}{N! (2R)^{2N(\frac{\Gamma}{4}-1)}}
  \int_{\mathbb{R}^{2N}} \prod_{1\leq j<k\leq N} |\tilde{z}_k-\tilde{z}_j|^{\Gamma} 
  \prod_{j=1}^{N} \frac{d^2\vec{\tilde{r}}_j
  }{\left(1+|\tilde{z}_j|^2\right)^{2+(N-1)\frac{\Gamma}{2}}}
\,.
\end{equation}
Let us set
\begin{equation}
Q_1=\frac{N}{N!} \int e^{-\beta U(z_1, \cdots,\, z_{N-1},\, 0)}
  dS_1\ldots dS_{N-1},  
\end{equation}
where the particle number $N$ is fixed at the north pole. By
definition, the density at the north pole is
\begin{equation}
  \rho_{(1)}(0)=\frac{Q_1}{Q}=\rho_b
  \,.
\end{equation}
In the coordinates of the stereographic projection of the sphere of
radius $R$, $Q_1$ can be written as
\begin{equation}
  \label{eq:Q1a}
  Q_1=\frac{N e^{\Gamma N^2/4}}{N! (2R)^{N^2\Gamma/2}}
  \int_{\mathbb{R}^{2(N-1)}} \prod_{1\leq j<k\leq N-1} |z_k-z_j|^{\Gamma} 
  \prod_{j=1}^{N-1} \frac{|z_j|^{\Gamma}\,d^2\vec{r}_j
  }{\left(1+\frac{|z_j|^2}{(2R)^2}\right)^{2+(N-1)\frac{\Gamma}{2}}}
\end{equation}
or, using the scaled coordinates $\tilde{z}$,
\begin{equation}
  \label{eq:Q1b}
  Q_1=\frac{N e^{\Gamma N^2/4}}{(2R)^{2(N\frac{\Gamma}{4}-N-1)}}
    \int_{\mathbb{R}^{2(N-1)}} \prod_{1\leq j<k\leq N-1} |\tilde{z}_k-\tilde{z}_j|^{\Gamma} 
    \prod_{j=1}^{N-1} \frac{|\tilde{z}_j|^{\Gamma}\,d^2\vec{\tilde{r}}_j
    }{\left(1+|\tilde{z}_j|^2\right)^{2+(N-1)\frac{\Gamma}{2}}}
\end{equation}
Computing the derivative of $Q_1$ with respect to $R$, using
expression~(\ref{eq:Q1a}) gives
\begin{equation}
  \label{eq:deriva}
  \frac{R}{Q}\frac{\partial Q_1}{\partial R}=
  -\frac{N^2\Gamma}{2} \rho_{(1)}(0) + 2
  \left(2+\frac{\Gamma}{2}(N-1)\right)
  \int \rho_{(2)}(\theta) \sin^2\frac{\theta}{2}\,dS.
\end{equation}
Here the last integral appears from the derivative
\begin{eqnarray}
R\frac{\partial}{\partial
  R}\left(\frac{1}{1+\frac{|z_j|^2}{(2R)^2}}\right)^{2+(N-1)\frac{\Gamma}{2}}=
2(2+(N-1)\frac{\Gamma}{2})
\frac{|z_j|^2/(2R)^2}{\left(1+\frac{|z_j|^2}{(2R)^2}\right)^{3+(N-1)\frac{\Gamma}{2}}}\\
=
2(2+(N-1)\frac{\Gamma}{2})
\left(\frac{1}{1+\frac{|z_j|^2}{(2R)^2}}\right)^{2+(N-1)\frac{\Gamma}{2}}
\sin^2\frac{\theta_j}{2}
\,.
\end{eqnarray}
On the other hand, starting with expression~(\ref{eq:Q1b}) gives
\begin{equation}
  \label{eq:derivb}
 \frac{R}{Q}\frac{\partial Q_1}{\partial R}=
2\left((N-1)-N\frac{\Gamma}{4}\right) \rho_{(1)}(0).
\end{equation}
Equating~(\ref{eq:deriva}) and~(\ref{eq:derivb}), using the fact
that $\rho_{(1)}(0)=\rho_b$, and subtracting on both sides $\int
\rho_b^2 \sin^2 (\theta/2) \,dS= N\rho_b/2$, gives
\begin{equation}
  \label{eq:sumrule}
  \int \left(2R\sin\frac{\theta}{2}\right)^2 \rho_{(2)}^T(\theta) \, dS=
  -  \frac{1}{\pi}\frac{N}{(N-1)\frac{\Gamma}{2}+2}
  \,,
\end{equation}
where $\rho_{(2)}^T(\theta)$ is the truncated pair correlation
function between the north pole and a point located on the sphere at
$(\theta,\phi)$.

In the planar limit $R\to\infty$, $N\to\infty$, with $\rho_b=N/(4\pi
R^2)$ finite, the length of the chord $r=2R\sin(\theta/2)$ becomes the
distance between the two points, and one recovers the
Stillinger-Lovett second moment sum rule~(\ref{1b}).

\subsection{Higher order moments}
\label{sec:higher-moments}

Let us define
\begin{equation}
  \hat I_{2n}=\rho_b \left(\frac{\pi \Gamma \rho_b}{2}\right)^n
  \int (2R\sin(\theta/2))^{2n} h(\theta)\, dS,
\end{equation}
where $h$ is the total correlation function
\begin{equation}
h(\theta)=\rho_{(2)}^T(\theta)/\rho_b^2  \,.
\end{equation}
Using the formalism presented in Section~\ref{S2} and~\cite{TF99}, the
density at a given point on the sphere is
\begin{equation}
\rho_{(1)}(\theta)=\frac{\rho_{b}((N-1)\frac{\Gamma}{2}+1)!
}{Z_{\sphere} N(1+x^2)^{(N-1)\Gamma/2}}
\sum_{\mu} \frac{(c_\mu^{(N)}(\Gamma/2))^2}{\prod_i m_i!} \prod_{l=1}^{N} \mu_l ! 
((N-1)\frac{\Gamma}{2}-\mu_l)! 
\sum_{k=1}^{N} \frac{x^{2\mu_k}}{\mu_k!((N-1)\frac{\Gamma}{2}-\mu_k)!} 
\end{equation}
where $\theta$ is the angle from the north pole of the sphere, 
$x=\tan(\theta/2)$, and
\begin{equation}
  Z_{\sphere}=\sum_{\mu} \frac{(c_\mu^{(N)}(\Gamma/2))^2}{\prod_i m_i!} \prod_{l=1}^{N} \mu_l ! 
((N-1)\frac{\Gamma}{2}-\mu_l)!
\,,
\end{equation}
is (up to a multiplicative constant) the partition function of the
2dOCP on the sphere. Since the sphere is homogeneous, the density does
not depend on $x$ and should be equal to $\rho_{b}$. This gives an
interesting relation satisfied by the coefficients,
$(c_\mu^{(N)}(\Gamma/2))^2$
\begin{equation}
  N (1+x)^{(N-1)\Gamma/2}= \frac{((N-1)\frac{\Gamma}{2}+1)!}{Z_{\sphere}}
  \sum_{\mu} \frac{(c_\mu^{(N)}(\Gamma/2))^2}{\prod_i m_i!} \prod_{l=1}^{N} \mu_l ! 
((N-1)\frac{\Gamma}{2}-\mu_l)! 
\sum_{k=1}^{N} \frac{x^{2\mu_k}}{\mu_k!((N-1)\frac{\Gamma}{2}-\mu_k)!} 
\end{equation}
Putting $x=0$, then relates a sum of all admissible
partitions and those with $\mu_N=0$,
\begin{equation}
  N  \sum_{\mu} \frac{(c_\mu^{(N)}(\Gamma/2))^2}{\prod_i m_i!} \prod_{l=1}^{N} \mu_l ! 
((N-1)\frac{\Gamma}{2}-\mu_l)! = ((N-1)\frac{\Gamma}{2}+1)!
  \sum_{\mu \atop \text{with }\mu_N=0} 
\frac{(c_\mu^{(N)}(\Gamma/2))^2}{\prod_i m_i!} \prod_{l=1}^{N-1} \mu_l ! 
((N-1)\frac{\Gamma}{2}-\mu_l)!
\end{equation}

The two-point correlation function is given by
\begin{equation}
  \label{eq:rho2}
  \rho_{(2)}(\theta)=\frac{\rho_b^2 [((N-1)\frac{\Gamma}{2}+1)!]^2
  }{Z_{\sphere} N^2 (1+x^2)^{(N-1)\Gamma/2}}
      \sum_{\mu \atop \text{with }\mu_N=0}
      \frac{(c_\mu^{(N)}(\Gamma/2))^2}{\prod_i m_i!} \prod_{l=1}^{N-1} \mu_l ! 
      ((N-1)\frac{\Gamma}{2}-\mu_l)!
      \sum_{k=1}^{N-1} \frac{x^{2\mu_k}}{\mu_k! ( (N-1)\frac{\Gamma}{2}-\mu_k)!}
\,.
\end{equation}
From these expressions, we obtain the $2n$-moment of the total
correlation function 
\begin{eqnarray}
  \label{eq:I2n}
  \hat I_{2n}&=&\left(\frac{N\Gamma}{2}\right)^2
      \Bigg[
        \frac{(((N-1)\frac{\Gamma}{2}+1)!)^2}{N((N-1)\frac{\Gamma}{2}+1+n)!
          Z_\sphere}
        \nonumber\\
        &&\times
        \sum_{\mu \atop \text{with }\mu_N=0} \frac{(c_\mu^{(N)}(\Gamma/2))^2
          \prod_{l=1}^{N-1} \mu_l!((N-1)\frac{\Gamma}{2}-\mu_l)!}{\prod_i
          m_{i}!}
    \sum_{k=1}^{N-1} \frac{(\mu_k+n)!}{\mu_k!}
    -\frac{N}{n+1}
    \Bigg]\,.
\end{eqnarray}
In the case $n=1$, the sum 
\begin{equation}
 \sum_{k=1}^{N-1}
 \frac{(\mu_k+n)!}{\mu_k!} =
 \sum_{k=1}^{N-1} (\mu_k+1)=(N-1)\left(\frac{\Gamma N}{4}+1\right)
 \,,
\end{equation}
simplifies and it is independent of the partition $\mu$, which lead to
the sum rule discussed in the previous section,
\begin{equation}
  \hat I_{2} = \frac{N\Gamma}{\Gamma-N\Gamma-4}
\end{equation}
Unfortunately, for $n\geq 2$, no such simplification seems
possible. 

For $\Gamma=2$, only one partition appears in the expansion
and the result for any $n$ is
\begin{equation}
  \hat I_{2n} = -N^{n} \frac{n! N!}{(N+n)!}\,,\qquad \Gamma=2
  \,.
\end{equation}
In the thermodynamic limit,
$N\to\infty$, the values of $\hat I_4$ and $\hat I_6$ are also
known for any $\Gamma$~\cite{KMST99}
\begin{eqnarray}
  \hat I_4 & = & \Gamma - 4 \,,\\
  \hat I_6 & = & \frac{3}{4}\, (\Gamma-6)(8-3\Gamma)\,.
\end{eqnarray}
For finite $N$, we computed the moments numerically
from~(\ref{eq:I2n}). Since the maximum value of $N$ for which the
numerical calculations where possible is small, the results for the
moments show important finite size corrections with respect to the
known values for $N\to\infty$. The exact results for $n=1$ (any
$\Gamma$), and the one for $\Gamma=2$, suggest that the finite size
corrections can be understood as a series expansion in powers of
$1/N$. Indeed, for $n=1$,
\begin{equation}
  \hat I_2 =-1 -\sum_{k=1}^{\infty} (-1)^k
  \left(\frac{4-\Gamma}{\Gamma}\right)^k
  \frac{1}{N^k}
\end{equation}
and for $\Gamma=2$,
\begin{eqnarray}
  \hat I_{2n}&=&-n! \Bigg[1 -\frac{n(n+1)}{2N} +
  \frac{n(n+1)(n+2)(1+3n)}{24 N^2}\nonumber\\
  &-&\frac{n^2(n+1)^2(n+2)(n+3)}{48 N^3} 
  +\frac{n(n+1)(n+2)(n+3)(n+4)(15n^3+30n^2+5n-2)}{5760 N^4}
  \nonumber\\
  &&+O(1/N^5)
  \Bigg]
\end{eqnarray}
Therefore, we decided to fit the numerical data obtained for the
moments with a $1/N$ expansion. The results are shown for $\Gamma=4$
in tables~\ref{tableGamma4I4}, \ref{tableGamma4I6},
\ref{tableGamma4I8}, for $\Gamma=6$ in tables \ref{tableGamma6I4},
\ref{tableGamma6I6}, \ref{tableGamma6I8}, and for $\Gamma=8$ in tables
\ref{tableGamma8I4}, \ref{tableGamma8I6}, \ref{tableGamma8I8}.  For
$\Gamma=4$, the coefficient of order 0 (which gives the value of $\hat
I_{2n}$ when $N\to\infty$), shows an acceptable convergence to the
expected value for $\hat I_4$ and $\hat I_6$, see
tables~\ref{tableGamma4I4}, \ref{tableGamma4I6}. Also, from
table~\ref{tableGamma4I8} ($\hat{I}_8$, $\Gamma=4$), the zero order
coefficient seems to converge to a value close to $-30$. This allows
us to obtain an estimate for the value of 8-th moment, for $\Gamma=4$,
in the thermodynamic limit, $\hat I_8\simeq -30$. Unfortunately, for
$\Gamma=6$ and $\Gamma=8$, since the maximum value of $N$ for which
the calculation were possible was smaller and the finite size
corrections still very large, we were not able to obtain any estimate
for $\hat I_8$.

\begin{table}[htbp]
  \begin{center}
\begin{tabular}{|r|r|r|r|r|r|}
\hline
\multicolumn{1}{|c|}{$N$} & \multicolumn{1}{c}{$\hat I_4 \qquad =$} &
\multicolumn{1}{c}{$\quad a \quad +$} & \multicolumn{1}{c}{$\quad b/N\quad +$} &
\multicolumn{1}{c}{$\quad c/N^2\quad +$} & \multicolumn{1}{c|}{$d/N^3$} \\ \hline
\multicolumn{1}{|l|}{} & \multicolumn{1}{l|}{} & \multicolumn{1}{c|}{$a$} & \multicolumn{1}{c|}{$b$} & \multicolumn{1}{c|}{$c$} & \multicolumn{1}{c|}{$d$} \\ \hline
2 & -1.06666666666667 & \multicolumn{1}{l|}{} & \multicolumn{1}{l|}{} & \multicolumn{1}{l|}{} & \multicolumn{1}{l|}{} \\ \hline
3 & -0.73469387755102 & \multicolumn{1}{l|}{} & \multicolumn{1}{l|}{} & \multicolumn{1}{l|}{} & \multicolumn{1}{l|}{} \\ \hline
4 & -0.552915766738661 & \multicolumn{1}{l|}{} & \multicolumn{1}{l|}{} & \multicolumn{1}{l|}{} & \multicolumn{1}{l|}{} \\ \hline
5 & -0.437781621713968 & 0.076709 & -2.81362 & 1.30721 & -0.506965 \\ \hline
6 & -0.361584090880502 & -0.042858 & -1.37881 & -4.31243 & 6.66705 \\ \hline
7 & -0.307439760694233 & 0.0121923 & -2.20457 & -0.238709 & 0.0610148 \\ \hline
8 & -0.267158562552772 & -0.00224307 & -1.94473 & -1.7833 & 3.09245 \\ \hline
9 & -0.236094785664912 & -0.00181613 & -1.9537 & -1.72096 & 2.94899 \\ \hline
10 & -0.211435346122364 & 0.0000393273 & -1.99823 & -1.36657 & 2.01384 \\ \hline
11 & -0.191399568920847 & 0.000150015 & -1.99312 & -1.41239 & 2.15017 \\ \hline
12 & -0.17480728582518 & -0.000241804 & -1.99036 & -1.43984 & 2.24104 \\ \hline
13 & -0.160846001528985 & -0.000138037 & -1.99379 & -1.40227 & 2.10407 \\ \hline
14 & -0.148938895568825 & -0.0000794746 & -1.99589 & -1.37703 & 2.00357 \\ \hline
\multicolumn{1}{|l|}{} & \multicolumn{1}{l|}{} & \multicolumn{1}{l|}{} & \multicolumn{1}{l|}{} & \multicolumn{1}{l|}{} & \multicolumn{1}{l|}{} \\ \hline
\multicolumn{1}{|l|}{$\infty$} & 0 & \multicolumn{1}{l|}{} & \multicolumn{1}{l|}{} & \multicolumn{1}{l|}{} & \multicolumn{1}{l|}{} \\ \hline
\end{tabular}
  \end{center}
\caption{$1/N$ expansion coefficients for $\hat I_4$ when $\Gamma=4$.}
\label{tableGamma4I4}
\end{table}

\begin{table}[htbp]
  \begin{center}
\begin{tabular}{|r|r|r|r|r|r|}
\hline
\multicolumn{1}{|c|}{$N$} & \multicolumn{1}{c}{$\hat I_6 \qquad =$} &
\multicolumn{1}{c}{$\quad a \quad +$} & \multicolumn{1}{c}{$\quad b/N\quad +$} &
\multicolumn{1}{c}{$\quad c/N^2\quad +$} & \multicolumn{1}{c|}{$d/N^3$} \\ \hline
\multicolumn{1}{|l|}{} & \multicolumn{1}{l|}{} & \multicolumn{1}{c|}{$a$} & \multicolumn{1}{c|}{$b$} & \multicolumn{1}{c|}{$c$} & \multicolumn{1}{c|}{$d$} \\ \hline
4 & 5.1605471562275 & \multicolumn{1}{l|}{} & \multicolumn{1}{l|}{} & \multicolumn{1}{l|}{} & \multicolumn{1}{l|}{} \\ \hline
5 & 6.11292630320115 & \multicolumn{1}{l|}{} & \multicolumn{1}{l|}{} & \multicolumn{1}{l|}{} & \multicolumn{1}{l|}{} \\ \hline
6 & 6.56602660174796 &  &  &  &  \\ \hline
7 & 6.78048920359037 & 5.49915 & 27.6653 & -150.595 & 138.066 \\ \hline
8 & 6.87498575554362 & 5.58979 & 26.0337 & -140.897 & 119.031 \\ \hline
9 & 6.9075123159669 & 5.74333 & 22.8094 & -118.48 & 67.4414 \\ \hline
10 & 6.90772192712509 & 5.87714 & 19.5982 & -92.9236 & 0.00502236 \\ \hline
11 & 6.89103601184945 & 5.92331 & 18.3514 & -81.7487 & -33.2425 \\ \hline
12 & 6.86574608931141 & 5.94863 & 17.592 & -74.1805 & -58.301 \\ \hline
13 & 6.83643498600859 & 5.96747 & 16.9702 & -67.3594 & -83.1735 \\ \hline
14 & 6.80567000027031 & 5.97905 & 16.5534 & -62.3686 & -103.044 \\ \hline
\multicolumn{1}{|l|}{} & \multicolumn{1}{l|}{} & \multicolumn{1}{l|}{} & \multicolumn{1}{l|}{} & \multicolumn{1}{l|}{} & \multicolumn{1}{l|}{} \\ \hline
\multicolumn{1}{|l|}{$\infty$} & 6 & \multicolumn{1}{l|}{} & \multicolumn{1}{l|}{} & \multicolumn{1}{l|}{} & \multicolumn{1}{l|}{} \\ \hline
\end{tabular}
  \end{center}
\caption{$1/N$ expansion coefficients for $\hat I_6$ when $\Gamma=4$.}
\label{tableGamma4I6}
\end{table}

\begin{table}[htbp]
  \begin{center}
\begin{tabular}{|r|r|l|l|l|l|}
\hline
\multicolumn{1}{|c|}{$N$} & \multicolumn{1}{c}{$\hat I_8 \qquad =$} &
\multicolumn{1}{c}{$\quad a \quad +$} & \multicolumn{1}{c}{$\quad b/N\quad +$} &
\multicolumn{1}{c}{$\quad c/N^2\quad +$} & \multicolumn{1}{c|}{$d/N^3$} \\ \hline
\multicolumn{1}{|l|}{} & \multicolumn{1}{l|}{} & \multicolumn{1}{c|}{$a$} & \multicolumn{1}{c|}{$b$} & \multicolumn{1}{c|}{$c$} & \multicolumn{1}{c|}{$d$} \\ \hline
6 & 40.3968590167648 &  &  &  &  \\ \hline
7 & 33.7964087943161 &  &  &  &  \\ \hline
8 & 26.9657510022336 &  &  &  &  \\ \hline
9 & 20.6758161117942 & \multicolumn{1}{r|}{-49.3234} & \multicolumn{1}{r|}{709.717} & \multicolumn{1}{r|}{-95.7946} & \multicolumn{1}{r|}{-5595.45} \\ \hline
10 & 15.1585480309705 & \multicolumn{1}{r|}{-38.7061} & \multicolumn{1}{r|}{454.9} & \multicolumn{1}{r|}{1932.12} & \multicolumn{1}{r|}{-10946.6} \\ \hline
11 & 10.414558865846 & \multicolumn{1}{r|}{-33.7202} & \multicolumn{1}{r|}{320.28} & \multicolumn{1}{r|}{3138.71} & \multicolumn{1}{r|}{-14536.5} \\ \hline
12 & 6.3618885422448 & \multicolumn{1}{r|}{-31.5027} & \multicolumn{1}{r|}{253.756} & \multicolumn{1}{r|}{3801.74} & \multicolumn{1}{r|}{-16731.8} \\ \hline
13 & 2.89953248501109 & \multicolumn{1}{r|}{-30.4953} & \multicolumn{1}{r|}{220.513} & \multicolumn{1}{r|}{4166.41} & \multicolumn{1}{r|}{-18061.5} \\ \hline
14 & -0.0685490396317987 & \multicolumn{1}{r|}{-30.1108} & \multicolumn{1}{r|}{206.671} & \multicolumn{1}{r|}{4332.12} & \multicolumn{1}{r|}{-18721.3} \\ \hline
\end{tabular}
  \end{center}
\caption{$1/N$ expansion coefficients for $\hat I_8$ when $\Gamma=4$.}
\label{tableGamma4I8}
\end{table}

\begin{table}[htbp]
  \begin{center}
\begin{tabular}{|r|r|l|l|l|l|}
\hline
\multicolumn{1}{|c|}{$N$} & \multicolumn{1}{c}{$\hat I_4 \qquad =$} &
\multicolumn{1}{c}{$\quad a \quad +$} & \multicolumn{1}{c}{$\quad b/N\quad +$} &
\multicolumn{1}{c}{$\quad c/N^2\quad +$} & \multicolumn{1}{c|}{$d/N^3$} \\ \hline
\multicolumn{1}{|l|}{} & \multicolumn{1}{l|}{} & \multicolumn{1}{c|}{$a$} & \multicolumn{1}{c|}{$b$} & \multicolumn{1}{c|}{$c$} & \multicolumn{1}{c|}{$d$} \\ \hline
4 & 1.77112299465241 &  &  &  &  \\ \hline
5 & 1.92727455514225 &  &  &  &  \\ \hline
6 & 1.97167595506035 &  &  &  &  \\ \hline
7 & 1.99917180579183 & \multicolumn{1}{r|}{2.90767} & \multicolumn{1}{r|}{-14.9074} & \multicolumn{1}{r|}{84.3562} & \multicolumn{1}{r|}{-171.645} \\ \hline
8 & 2.01319334170647 & \multicolumn{1}{r|}{1.72398} & \multicolumn{1}{r|}{6.39891} & \multicolumn{1}{r|}{-42.2979} & \multicolumn{1}{r|}{76.9278} \\ \hline
9 & 2.02020128573558 & \multicolumn{1}{r|}{1.95459} & \multicolumn{1}{r|}{1.55616} & \multicolumn{1}{r|}{-8.62918} & \multicolumn{1}{r|}{-0.556218} \\ \hline
10 & 2.02414460391682 & \multicolumn{1}{r|}{2.08557} & \multicolumn{1}{r|}{-1.5874} & \multicolumn{1}{r|}{16.3883} & \multicolumn{1}{r|}{-66.5709} \\ \hline
11 & 2.02625080279343 & \multicolumn{1}{r|}{1.98854} & \multicolumn{1}{r|}{1.03254} & \multicolumn{1}{r|}{-7.09409} & \multicolumn{1}{r|}{3.29402} \\ \hline
12 & 2.02719976576841 & \multicolumn{1}{r|}{1.98147} & \multicolumn{1}{r|}{1.2446} & \multicolumn{1}{r|}{-9.2077} & \multicolumn{1}{r|}{10.2923} \\ \hline
\multicolumn{1}{|l|}{} & \multicolumn{1}{l|}{} &  &  &  &  \\ \hline
\multicolumn{1}{|l|}{$\infty$} & 2 &  &  &  &  \\ \hline
\end{tabular}
  \end{center}
\caption{$1/N$ expansion coefficients for $\hat I_4$ when $\Gamma=6$.}
\label{tableGamma6I4}
\end{table}

\begin{table}[htbp]
  \begin{center}
\begin{tabular}{|r|r|l|l|l|l|}
\hline
\multicolumn{1}{|c|}{$N$} & \multicolumn{1}{c}{$\hat I_6 \qquad =$} &
\multicolumn{1}{c}{$\quad a \quad +$} & \multicolumn{1}{c}{$\quad b/N\quad +$} &
\multicolumn{1}{c}{$\quad c/N^2\quad +$} & \multicolumn{1}{c|}{$d/N^3$} \\ \hline
\multicolumn{1}{|l|}{} & \multicolumn{1}{l|}{} & \multicolumn{1}{c|}{$a$} & \multicolumn{1}{c|}{$b$} & \multicolumn{1}{c|}{$c$} & \multicolumn{1}{c|}{$d$} \\ \hline
4 & 29.2857260386672 &  &  &  &  \\ \hline
5 & 24.3553395496129 &  &  &  &  \\ \hline
6 & 19.3570221116088 &  &  &  &  \\ \hline
7 & 15.8157978386411 & \multicolumn{1}{r|}{23.4057} & \multicolumn{1}{r|}{-367.801} & \multicolumn{1}{r|}{3052.6} & \multicolumn{1}{r|}{-5949.24} \\ \hline
8 & 13.2137920826172 & \multicolumn{1}{r|}{-1.67692} & \multicolumn{1}{r|}{83.6859} & \multicolumn{1}{r|}{368.754} & \multicolumn{1}{r|}{-681.887} \\ \hline
9 & 11.2455089383337 & \multicolumn{1}{r|}{-0.844904} & \multicolumn{1}{r|}{66.2135} & \multicolumn{1}{r|}{490.229} & \multicolumn{1}{r|}{-961.446} \\ \hline
10 & 9.75173060605997 & \multicolumn{1}{r|}{4.89476} & \multicolumn{1}{r|}{-71.5384} & \multicolumn{1}{r|}{1586.5} & \multicolumn{1}{r|}{-3854.23} \\ \hline
11 & 8.58562698480205 & \multicolumn{1}{r|}{0.124033} & \multicolumn{1}{r|}{57.2712} & \multicolumn{1}{r|}{431.99} & \multicolumn{1}{r|}{-419.314} \\ \hline
12 & 7.65247243254821 & \multicolumn{1}{r|}{-0.286731} & \multicolumn{1}{r|}{69.5941} & \multicolumn{1}{r|}{309.171} & \multicolumn{1}{r|}{-12.6574} \\ \hline
\multicolumn{1}{|l|}{} & \multicolumn{1}{l|}{} &  &  &  &  \\ \hline
\multicolumn{1}{|l|}{$\infty$} & 0 &  &  &  &  \\ \hline
\end{tabular}
  \end{center}
\caption{$1/N$ expansion coefficients for $\hat I_6$ when $\Gamma=6$.}
\label{tableGamma6I6}
\end{table}

\begin{table}[htbp]
  \begin{center}
\begin{tabular}{|r|r|l|l|l|l|}
\hline
\multicolumn{1}{|c|}{$N$} & \multicolumn{1}{c}{$\hat I_8 \qquad =$} &
\multicolumn{1}{c}{$\quad a \quad +$} & \multicolumn{1}{c}{$\quad b/N\quad +$} &
\multicolumn{1}{c}{$\quad c/N^2\quad +$} & \multicolumn{1}{c|}{$d/N^3$} \\ \hline
\multicolumn{1}{|l|}{} & \multicolumn{1}{l|}{} & \multicolumn{1}{c|}{$a$} & \multicolumn{1}{c|}{$b$} & \multicolumn{1}{c|}{$c$} & \multicolumn{1}{c|}{$d$} \\ \hline
4 & 157.436163836164 &  &  &  &  \\ \hline
5 & -33.5270610707337 &  &  &  &  \\ \hline
6 & -196.111013214468 &  &  &  &  \\ \hline
7 & -293.318966661244 &  &  &  &  \\ \hline
8 & -343.529970425059 &  &  &  &  \\ \hline
9 & -366.832915368496 & \multicolumn{1}{r|}{129.267} & \multicolumn{1}{r|}{-11229.9} & \multicolumn{1}{r|}{71322.9} & \multicolumn{1}{r|}{-93943.9} \\ \hline
10 & -374.243763793112 & \multicolumn{1}{r|}{161.032} & \multicolumn{1}{r|}{-11992.2} & \multicolumn{1}{r|}{77390} & \multicolumn{1}{r|}{-109954} \\ \hline
11 & -373.310338904328 & \multicolumn{1}{r|}{-86.8351} & \multicolumn{1}{r|}{-5299.81} & \multicolumn{1}{r|}{17406.2} & \multicolumn{1}{r|}{68510.9} \\ \hline
12 & -368.298029500803 & \multicolumn{1}{r|}{-183.485} & \multicolumn{1}{r|}{-2400.33} & \multicolumn{1}{r|}{-11492.1} & \multicolumn{1}{r|}{64194} \\ \hline
\end{tabular}
  \end{center}
\caption{$1/N$ expansion coefficients for $\hat I_8$ when $\Gamma=6$.}
\label{tableGamma6I8}
\end{table}

\begin{table}[htbp]
  \begin{center}
\begin{tabular}{|r|r|l|l|l|l|}
\hline
\multicolumn{1}{|c|}{$N$} & \multicolumn{1}{c}{$\hat I_4 \qquad =$} &
\multicolumn{1}{c}{$\quad a \quad +$} & \multicolumn{1}{c}{$\quad b/N\quad +$} &
\multicolumn{1}{c}{$\quad c/N^2\quad +$} & \multicolumn{1}{c|}{$d/N^3$} \\ \hline
\multicolumn{1}{|l|}{} & \multicolumn{1}{l|}{} & \multicolumn{1}{c|}{$a$} & \multicolumn{1}{c|}{$b$} & \multicolumn{1}{c|}{$c$} & \multicolumn{1}{c|}{$d$} \\ \hline
3 & 5.9504132231405 &  &  &  &  \\ \hline
4 & 5.35216744227873 &  &  &  &  \\ \hline
5 & 5.31949584221798 &  &  &  &  \\ \hline
6 & 5.03608123946663 & \multicolumn{1}{r|}{-6.67707} & \multicolumn{1}{r|}{150.385} & \multicolumn{1}{r|}{-623.767} & \multicolumn{1}{r|}{858.777} \\ \hline
7 & 4.90673278031246 & \multicolumn{1}{r|}{11.9835} & \multicolumn{1}{r|}{-129.523} & \multicolumn{1}{r|}{757.114} & \multicolumn{1}{r|}{-1380.49} \\ \hline
\multicolumn{1}{|l|}{} & \multicolumn{1}{l|}{} &  &  &  &  \\ \hline
\multicolumn{1}{|l|}{$\infty$} & 4 &  &  &  &  \\ \hline
\end{tabular}
  \end{center}
\caption{$1/N$ expansion coefficients for $\hat I_4$ when $\Gamma=8$.}
\label{tableGamma8I4}
\end{table}

\begin{table}[htbp]
  \begin{center}
\begin{tabular}{|r|r|l|l|l|l|}
\hline
\multicolumn{1}{|c|}{$N$} & \multicolumn{1}{c}{$\hat I_6 \qquad =$} &
\multicolumn{1}{c}{$\quad a \quad +$} & \multicolumn{1}{c}{$\quad b/N\quad +$} &
\multicolumn{1}{c}{$\quad c/N^2\quad +$} & \multicolumn{1}{c|}{$d/N^3$} \\ \hline
\multicolumn{1}{|l|}{} & \multicolumn{1}{l|}{} & \multicolumn{1}{c|}{$a$} & \multicolumn{1}{c|}{$b$} & \multicolumn{1}{c|}{$c$} & \multicolumn{1}{c|}{$d$} \\ \hline
3 & 103.537190082645 &  &  &  &  \\ \hline
4 & 57.4074000134889 &  &  &  &  \\ \hline
5 & 30.083006000112 &  &  &  &  \\ \hline
6 & 8.52435062302512 & \multicolumn{1}{r|}{-202.192} & \multicolumn{1}{r|}{2030.17} & \multicolumn{1}{r|}{-5851.51} & \multicolumn{1}{r|}{7537.7} \\ \hline
7 & -1.73452473741117 & \multicolumn{1}{r|}{248.055} & \multicolumn{1}{r|}{-4723.54} & \multicolumn{1}{r|}{27466.8} & \multicolumn{1}{r|}{-4649} \\ \hline
\multicolumn{1}{|l|}{} & \multicolumn{1}{l|}{} &  &  &  &  \\ \hline
\multicolumn{1}{|l|}{$\infty$} & -24 &  &  &  &  \\ \hline
\end{tabular}
      \end{center}
\caption{$1/N$ expansion coefficients for $\hat I_6$ when $\Gamma=8$.}
\label{tableGamma8I6}
\end{table}

\begin{table}[htbp]
\begin{center}
\begin{tabular}{|r|r|l|l|l|l|}
\hline
\multicolumn{1}{|c|}{$N$} & \multicolumn{1}{c}{$\hat I_8 \qquad =$} &
\multicolumn{1}{c}{$\quad a \quad +$} & \multicolumn{1}{c}{$\quad b/N\quad +$} &
\multicolumn{1}{c}{$\quad c/N^2\quad +$} & \multicolumn{1}{c|}{$d/N^3$} \\ \hline
\multicolumn{1}{|l|}{} & \multicolumn{1}{l|}{} & \multicolumn{1}{c|}{$a$} & \multicolumn{1}{c|}{$b$} & \multicolumn{1}{c|}{$c$} & \multicolumn{1}{c|}{$d$} \\ \hline
3 & 1073.4573622182 &  &  &  &  \\ \hline
4 & -228.646302651363 &  &  &  &  \\ \hline
5 & -1293.26989559543 &  &  &  &  \\ \hline
6 & -1868.87943727765 & \multicolumn{1}{r|}{1402.47} & \multicolumn{1}{r|}{-68534} & \multicolumn{1}{r|}{384230} & \multicolumn{1}{r|}{-544767} \\ \hline
7 & -1927.54995720038 & \multicolumn{1}{r|}{13256.9} & \multicolumn{1}{r|}{246350} & $1.26146\times10^6$ & $-1.9673\times10^6$ \\ \hline
\end{tabular}
\end{center}
\caption{$1/N$ expansion coefficients for $\hat I_8$ when $\Gamma=8$.}
\label{tableGamma8I8}
\end{table}

%%%%%%%%%%%%%%%%%%%%%%%%%%%%%%%%%%%%%%%%%%%%%
\break  % to avoid tables from previous section overlapping in the
        % following section

\section{Diagrammatic expansions}
\label{S4}

The proof of the sixth moment sum rule~\cite{KMST99} is based on an
analysis of the Mayer diagrammatic expansion of the direct correlation
function, in the flat space in the bulk. Therefore, it is interesting
to study this expansion on the sphere. 

The direct correlation function $c$ is defined by the Ornstein-Zernike
equation
\begin{equation}
  \label{eq:OZ}
  h(\r_1,\r_2)=c(\r_1,\r_2) + \rho \int h(\r_1,\r_3) c(\r_3,\r_2) dS_3,
\end{equation}
where $\r_{i}$ points the direction of a point in the sphere from its
center. Since the system is homogeneous, both $h(\r_1,\r_2)$ and
$c(\r_1,\r_2)$ depend only the angle $\theta_{12}=(\r_1,\r_2)$. It is
useful to introduce the expansion of $h$ and $c$ in Legendre
polynomials
\begin{equation}
  h(\theta)=\sum_{l=0}^{\infty} h_l P_l(\cos \theta)
\end{equation}
\begin{equation}
  h_l=\frac{2l+1}{2}\int_{-1}^{1} h(x) P_{l}(x)\,dx
\end{equation}
and similarly for $c$. The Ornstein-Zernike equation reads
\begin{equation}
  \label{eq:OZ-fourier}
  h_l=c_l + \frac{N}{2l+1} h_l c_l
  \,.
\end{equation}
implying
\begin{equation}
  \label{eq:hl}
  h_{l}=\frac{c_l}{\displaystyle 1-\frac{N}{2l+1} c_{l}}
  \,.
\end{equation}
Notice also that the moments $\hat{I}_{2n}$ of the total correlation function are
directly related to the coefficients $h_l$, since $P_l(\cos\theta)$
can be expressed as a polynomial in $\sin^2(\theta/2)$ using the formula
$\cos\theta = 1- 2\sin^{2} (\theta/2)$.

As explained in detail in~\cite{KMST99}, the direct correlation
function has a renormalized diagrammatic expansion which reads
\begin{equation}
  \label{eq:c-direct}
  c(\r_1,\r_2)=-\Gamma v(\r_1,\r_2) + c^{(0)}(\r_1,\r_2) 
+\sum_{s=1}^{\infty} c^{(s)}(\r_1,\r_2)
\end{equation}
where $v(\r_1,\r_2)$ is the Coulomb potential between to unit
charges. $c^{(0)}$ is the renormalized ``watermelon'' Meeron graph
\begin{equation}
  c^{(0)}(\r_1,\r_2)=\frac{1}{2} K^2(\r_1,\r_2)
\end{equation}
where $K$ is defined by the equation
\begin{equation}
  \label{eq:K}
  K(\r_1,\r_2)=- \Gamma v(\r_1,\r_2) -  \rho \int  \Gamma
  v(\r_1,\r_3)K(\r_3,\r_1) dS_3
  \,.
\end{equation}
The rest of the expansion of $c$, $\sum_{s=1}^{\infty}
c^{(s)}(\r_1,\r_2)$ correspond to the remaining diagrams;
see~\cite{KMST99} for details.

The coefficients $K_l$ of the expansion of $K$ in Legendre polynomials
can be obtained from~(\ref{eq:K})
\begin{equation}
  K_{l}=\frac{-\Gamma v_l}{\displaystyle 1+\frac{N\Gamma v_l}{2l+1}},
\end{equation}
where the coefficients of the Coulomb potential are obtained by
solving Poisson equation in the sphere
\begin{equation}
  v_{l}=\frac{2l+1}{2l(l+1)}\,, \quad l>0.
\end{equation}
Notice that the coefficient for $l=0$ is not defined properly,
rather as shown in \cite{Tellez-DHsphere} it is a constant that can be chosen arbitrary.
Then
\begin{equation}
  K_l=\frac{-\Gamma (2l+1)}{2l(l+1)+N\Gamma}
  \,,\quad l>0\,.
\end{equation}

In~\cite{KMST99} it is proved that the zeroth and second moment of each
$c^{(s)}$ vanishes, therefore they do not contribute to small-$k$
expansion of the Fourier transform of $c$ up to terms of order
$k^{4}$. Then the structure of the Ornstein-Zernike equation in
Fourier space guaranties that these terms do not contribute to the
expansion of the total correlation function $h$ up to $k^8$. This
fixes the value of the moments up to the sixth. Unfortunately, on the
sphere this argument breaks down. If one can show that the zeroth and
second moment of $c^{(s)}$ also vanish on the sphere, this fixes the
values of $c_{0}$, and $c_{1}$, and from
eq.~(\ref{eq:OZ-fourier}) one can deduce the value of $h_0$,
$h_1$ and therefore $\hat{I}_{0}$ and $\hat{I}_{2}$ but unfortunately
one cannot deduce anything exact regarding $h_{2}$, $h_{3}$, and
$\hat{I}_4$, $\hat I_6$.

\subsection{Second moment}

Suppose, as in the plane, that $c^{(s)}$ has vanishing zeroth and second
moment, i.e.~$c^{(s)}_{0}=0$, and $c^{(s)}_{1}=0$. Then to compute
$c_{l}$, eq.~(\ref{eq:c-direct}) with the sum ignored implies
\begin{equation}
  \label{eq:cl}
  c_{l}=-\Gamma v_l + c_{l}^{(0)}=-\Gamma v_l + \frac{1}{2} m_{l}\,.
\end{equation}
The coefficients $c_{l}^{(0)}=m_l/2$ are obtained from those of the
expansion of $K^2$,
\begin{equation}
  K^2(\theta)=\sum_{l=0}^{\infty} m_l P_l(\cos\theta)
\end{equation}
which can be obtained if the expansion of a product of two Legendre
functions is known. If
\begin{equation}
  P_{l}(x)  P_{l'}(x) = \sum_{l''=0}^{\infty} p^{ll'}_{l''} P_{l''}(x)
\end{equation}
with
\begin{equation}
  \label{eq:pl}
  p_{l''}^{ll'}=\frac{2l''+1}{2}\int_{-1}^{1} P_{l''}(x) P_{l'}(x)
  P_{l}(x) \,dx
\end{equation}
then
\begin{equation}
  m_{l''}=\sum_{l=0}^{\infty}\sum_{l'=0}^{\infty} K_{l} K_{l'}
  p_{l''}^{ll'}
  \,.
\end{equation}
The $p_{l''}^{ll'}$ can be computed~\cite{Jackson} using the recursion
equation
\begin{equation}
  \label{eq:rec}
  (2l+1)xP_{l}(x)=l P_{l-1}(x) +(l+1) P_{l+1}(x)
  \,.
\end{equation}
For instance, for $l''=1$, using $x=P_{1}(x)$ in~(\ref{eq:rec}) and
replacing it in~(\ref{eq:pl}) gives~\cite{Jackson}
\begin{equation}
  p_{1}^{ll'}=\frac{3}{2}
  \left[\frac{2(l+1)}{(2l+1)(2l+3)}\delta_{l',l+1}+
    \frac{2l}{(2l-1)(2l+1)}\delta_{l',l-1}\right].
\end{equation}
Therefore,
\begin{eqnarray}
  \label{eq:m1-c}
  m_{1}&=&3\sum_{l=0}^{\infty} \frac{2(l+1)}{(2l+1)(2l+3)}\,K_{l+1}
  K_{l}
  \nonumber\\
  &=& 6 \Gamma^2 \sum_{l=0}^{\infty}
  \frac{l+1}{(2(l+1)l+N\Gamma)(2(l+1)(l+2)+N\Gamma)} 
 \nonumber\\
 &=&  \frac{3 \Gamma^2 }{2} \sum_{l=0}^{\infty}\left[
 \frac{1}{2l(l+1)+N\Gamma} - 
 \frac{1}{2(l+1)(l+2)+N\Gamma}\right]
 \nonumber\\
 &=& \frac{3\Gamma}{2N}.
 \end{eqnarray}
Then, replacing~(\ref{eq:m1-c}), into~(\ref{eq:cl}) and~(\ref{eq:hl})
\begin{equation}
  \label{eq:h1res}
  h_{1}= \frac{\displaystyle -\frac{3\Gamma}{4}\left(1-\frac{1}{N}\right)}%
{\displaystyle 1+\frac{N\Gamma}{4}-\frac{\Gamma}{4}}
\,.
\end{equation}
But
\begin{eqnarray}
  h_1&=&\frac{3}{2} \int h(\theta) (1-2\sin^2(\theta/2))\, d(\cos\theta)
  \nonumber\\
  &=&
  \frac{3}{N}\left( \hat{I}_0 - \frac{4}{N\Gamma} \hat{I}_2 \right)
  \,.
\end{eqnarray}
Since $\hat{I}_0=-1$, we verify that (\ref{eq:h1res}) is compatible
with the sum rule from section~\ref{S3},
$\hat{I}_2=N\Gamma/(\Gamma-N\Gamma-4)$.

Since the second moment sum rule was proven on firm grounds in
section~\ref{S3}, going backwards, this proves indirectly
that $\sum_{s=1}^{\infty} c_{1}^{(s)}=0$.

\subsection{Higher moments}

Although there is no reason to suppose that the contributions from
$\sum_{s=1}^{\infty} c_{l}^{(s)}$ for $l=2$ and $l=3$ will vanish for
a finite sphere, it is interesting to compute the contribution from
$-\Gamma v_{l} + c^{(0)}$ to the 4th and 6th moments. Let us define an
approximate direct correlation function $c^{\ap}$ by
\begin{equation}
  c^{\ap}(\theta)=\sum_{l=0}^{\infty} c_{l}^{\ap} P_l(\cos\theta)
\end{equation}
with
\begin{equation}
  \label{eq:clapprox}
  c_{l}^{\ap}=-\Gamma v_l + \frac{1}{2} m_{l}\,,
\end{equation}
and the corresponding total correlation function $h^{\ap}$
obtained from Ornstein-Zernike equation
\begin{equation}
  h^{\ap}_{l}=\frac{
    \displaystyle \frac{-(2l+1)\Gamma}{2l(l+1)}+\frac{m_l}{2}}{
    \displaystyle 1+\frac{N\Gamma}{2l(l+1)}-\frac{N m_l}{2(2l+1)}}
  \,.
\end{equation}

\subsection{Fourth moment}

The fourth moment is related to the coefficient $h_2$ by 
\begin{eqnarray}
  h_2&=&\frac{5\rho_b}{N}\int h(\theta) \left[
    1-\frac{3}{2}(2\sin(\theta/2))^2+\frac{3}{8}(2\sin(\theta/2))^4
    \right]\,dS
  \nonumber\\
  &=&
  \frac{5}{N}\left[ \hat{I}_0 -\frac{12}{N\Gamma} \hat{I}_2
    +\frac{24}{(N\Gamma)^2} \hat{I}_4\right]
\,,
\end{eqnarray}
and so
\begin{equation}
  \label{eq:I4h2}
  \hat{I}_4=
  \frac{(N\Gamma)^2}{24}
  \left[\frac{Nh_2}{5}+1+\frac{12}{\Gamma-N\Gamma-4}
    \right]\,.
\end{equation}
The quantity $h^{\ap}_{2}$ is given by
\begin{equation}
  \label{eq:h2app}
  h^{\ap}_2=\frac{-\frac{5\Gamma}{12}+\frac{m_2}{2}}{
    1+\frac{N\Gamma}{12}-\frac{Nm_2}{10}}
\end{equation}
with
\begin{equation}
  m_2=\sum_{l=0}^{\infty}\sum_{l'=0}^{\infty} K_l K_{l'} p_{2}^{ll'}
\end{equation}
and~\cite{Jackson}
\begin{equation}
  p_{2}^{ll'}=
  \frac{5}{4}\left[
    \frac{6 (l+1)(l+2)}{(2l+1)(2l+3)(2l+5)}\,\delta_{l',l+2}+
    \frac{6 (l'+1)(l'+2)}{(2l'+1)(2l'+3)(2l'+5)}\,\delta_{l,l'+2}
    -\frac{2}{2l+1}\,\delta_{ll'}\right].
\end{equation}
Then
\begin{eqnarray}
  m_2&=&\sum_{l=0}^{\infty}
  \frac{15 K_l K_{l+1} (l+1)(l+2)}{(2l+1)(2l+3)(2l+5)}
  \nonumber\\
  &&+ \sum_{l=0}^{\infty} \frac{5K_{l}^2 l (l+1)}{(2l-1)(2l+1)(2l+3)}
 \nonumber\\
 &=& \Gamma^2 f(N\Gamma)  
\end{eqnarray}
with
\begin{equation}
  f(x)=\sum_{l=0}^{\infty}\left[
  \frac{15 (l+1)(l+2)}{(2l(l+1)+x)(2(l+2)(l+3)+x)(2l+3)}
    +\frac{5(2l+1)l(l+1)}{(2l(l+1)+x)^2(2l-1)(2l+3)}
    \right].
\end{equation}
Replacing this into~(\ref{eq:h2app}) and (\ref{eq:I4h2}) gives an
approximation for the fourth moment $\hat{I}_{4}^{\ap}$.

\subsection{Sixth moment}

Following similar steps, one can obtain an approximation for
$\hat{I}_6$. The sixth moment is related to $h_3$ by
\begin{equation}
  h_{3}=\frac{7}{N}\left[ \hat{I}_0
    -\frac{24\hat{I}_2}{N\Gamma}
    +\frac{120 \hat{I}_4}{(N\Gamma)^2}
    -\frac{160 \hat{I}_6}{(N\Gamma)^3}\right]
\end{equation}
Therefore, using~(\ref{eq:I4h2}) and the known values of $\hat{I}_0$
and $\hat{I}_2$,
\begin{equation}
  \label{eq:I6h3}
  \hat{I}_6=(N\Gamma)^3 \left[
    -\frac{Nh_3}{1120}+\frac{Nh_2}{160}+\frac{1}{40}
    +\frac{9}{40(\Gamma-N\Gamma-4)}
    \right]
\end{equation}
$h_2$ can be approximated by $h_2^{\ap}$ from last section, and $h_3$
can be approximated by
\begin{equation}
  \label{eq:h3app}
  h^{\ap}_3=\frac{-\frac{7\Gamma}{24}+\frac{m_3}{2}}{
    1+\frac{N\Gamma}{24}-\frac{Nm_3}{14}}
\end{equation}
with
\begin{equation}
  \label{eq:m3def}
 m_3=\sum_{l=0}^{\infty} \sum_{l'=0}^{\infty} K_l K_{l'} p_3^{ll'}
\end{equation}
and
\begin{equation}
  \label{eq:p3def}
  p_{3}^{ll'}=\frac{7}{4}\int_{-1}^{1} (5x^3-3x) P_l(x) P_{l'}(x)\,dx
  \,.
\end{equation}
$p_{3}^{ll'}$ can be computed by applying the recursion equation~(\ref{eq:rec})
several times to obtain
\begin{eqnarray}
  x^3 P_{l}(x)&=&
\frac{1}{2l+1}
\left[\frac{l(l-1)(l-2)}{(2l-1)(2l-3)} P_{l-3}(x)
+\frac{3l(l^2-2)}{(2l-3)(2l+3)}P_{l-1}(x)
\right.
\nonumber\\
&&
\left.
+\frac{(l+1)(l+2)(l+3)}{(2l+3)(2l+5)}P_{l+3}(x)
+\frac{3(l+1)((l+1)^2-2)}{(2l-1)(2l+5)} P_{l+1}(x)
\right].
\nonumber\\
\end{eqnarray}
Then replacing into~(\ref{eq:p3def})
\begin{eqnarray}
  \label{eq:p3}
  p_{3}^{ll'}&=&\frac{7}{2(2l+1)}
  \left[\frac{5l(l-1)(l-2)}{(2l-1)(2l-3)(2l-5)}\,\delta_{l',l-3}
    +\frac{3l(l^2-1)}{(2l-3)(2l+3)(2l-1)}\,\delta_{l',l-1}
    \right.
    \nonumber\\
    &&    \left.
    +\frac{5(l+1)(l+2)(l+3)}{(2l+3)(2l+5)(2l+7)}\,\delta_{l',l+3}
    +\frac{3(l+1)((l+1)^2-1)}{(2l-1)(2l+5)(2l+3)}\,\delta_{l',l+1}
    \right]\,.
  \nonumber\\
\end{eqnarray}
Replacing this result into~(\ref{eq:m3def}) one obtains
\begin{eqnarray}
  m_3&=&7\Gamma^2
  \sum_{l=0}^{\infty}
  \frac{(l+1)(l+2)}{(2l+5)(2l(l+1)+N\Gamma)}
  \left[
    \frac{3l}{(2(l+1)(l+2)+N\Gamma)(2l-1)}
    \right.
    \nonumber\\
    &&
    \left.
    +\frac{5(l+3)}{(2(l+3)(l+4)+N\Gamma)(2l+3)}
    \right]
  \nonumber\\
  &=&
  \frac{7(2+3N\Gamma)\Gamma}{2N(12+3N\Gamma)}
  \,.
\end{eqnarray}
Then
\begin{equation}
  h_{3}^{\ap}=
  \frac{-7(-4+N(4-6\Gamma)+N^2\Gamma)\Gamma}{
    N(96+4(7N-1)\Gamma+(6-N)N\Gamma^2)}
  \,.
\end{equation}
$h_{3}^{\ap}$ and $h_{2}^{\ap}$ can be replaced
into~(\ref{eq:I6h3}) to obtain an approximation for the sixth moment
$\hat{I}_6^{\ap}$.

\subsection{Comparison with numerical results}

The expressions for the fourth and sixth moments can be compared to
their exact values which have been computed numerically in
section~\ref{sec:higher-moments}. Tables~\ref{tab:G4Idiag},
\ref{tab:G6Idiag}, \ref{tab:G8Idiag}, \ref{tab:G2Idiag} show the exact
value of the moments and the approximate value obtained from the
diagrammatic technique. For the fourth moment $\hat{I}_4^{\ap}$ gives
an acceptable approximation, on the other hand $\hat{I}_6^{\ap}$ has
accumulated too many errors from $h_{2}^{\ap}$ and $h_{3}^{\ap}$ and
gives a poor agreement with the exact value. However, as $N\to\infty$
both $\hat{I}_{4}^{\ap}$ and $\hat{I}_{6}^{\ap}$ approach the known
thermodynamic limit, as expected.

\begin{table}[htbp]
\begin{tabular}{|r|r|r|r|r|r|r|}
\hline
\multicolumn{1}{|l|}{$N$} & \multicolumn{1}{l|}{$\hat{I}_4$} & \multicolumn{1}{l|}{$\hat{I}_4^{\ap}$} & \multicolumn{1}{l|}{$\hat{I}_4$ error} & \multicolumn{1}{l|}{$\hat{I}_6$} & \multicolumn{1}{l|}{$\hat{I}_6^{\ap}$} & \multicolumn{1}{l|}{$\hat{I}_6$ error} \\ \hline
\multicolumn{1}{|l|}{} & \multicolumn{1}{l|}{} & \multicolumn{1}{l|}{} & \multicolumn{1}{l|}{} & \multicolumn{1}{l|}{} & \multicolumn{1}{l|}{} & \multicolumn{1}{l|}{} \\ \hline
2 & -1.06666666666667 & -0.686993 & 35.6\% & 0 & 0.241677 & \multicolumn{1}{l|}{} \\ \hline
3 & -0.73469387755102 & -0.534741 & 27.2\% & 3.30612244897959 & 1.54027 & 53.4\% \\ \hline
4 & -0.552915766738661 & -0.42966 & 22.3\% & 5.1605471562275 & 2.52407 & 51.1\% \\ \hline
5 & -0.437781621713968 & -0.356903 & 18.5\% & 6.11292630320115 & 3.21788 & 47.4\% \\ \hline
6 & -0.361584090880502 & -0.304426 & 15.8\% & 6.56602660174796 & 3.71182 & 43.5\% \\ \hline
7 & -0.307439760694233 & -0.26506 & 13.8\% & 6.78048920359037 & 4.07309 & 39.9\% \\ \hline
8 & -0.267158562552772 & -0.234542 & 12.2\% & 6.87498575554362 & 4.34502 & 36.8\% \\ \hline
9 & -0.236094785664912 & -0.210236 & 11.0\% & 6.9075123159669 & 4.5552 & 34.1\% \\ \hline
10 & -0.211435346122364 & -0.190444 & 9.9\% & 6.90772192712509 & 4.72147 & 31.6\% \\ \hline
11 & -0.191399568920847 & -0.174026 & 9.1\% & 6.89103601184945 & 4.85568 & 29.5\% \\ \hline
12 & -0.17480728582518 & -0.160195 & 8.4\% & 6.86574608931141 & 4.96591 & 27.7\% \\ \hline
13 & -0.160846001528985 & -0.148387 & 7.7\% & 6.83643498600859 & 5.05783 & 26.0\% \\ \hline
14 & -0.148938895568825 & -0.138191 & 7.2\% & 6.80567000027031 & 5.1355 & 24.5\% \\ \hline
\multicolumn{1}{|l|}{} & \multicolumn{1}{l|}{} & \multicolumn{1}{l|}{} & \multicolumn{1}{l|}{} & \multicolumn{1}{l|}{} & \multicolumn{1}{l|}{} & \multicolumn{1}{l|}{} \\ \hline
\multicolumn{1}{|l|}{$\infty$} & 0 & \multicolumn{1}{l|}{} & \multicolumn{1}{l|}{} & 6 & \multicolumn{1}{l|}{} & \multicolumn{1}{l|}{} \\ \hline
\end{tabular}
\caption{Exact value of the moments and their approximation from the diagrammatic technique for $\Gamma=4$.}
\label{tab:G4Idiag}
\end{table}

\begin{table}[htbp]
\begin{tabular}{|r|r|r|r|r|r|r|}
\hline
\multicolumn{1}{|l|}{$N$} & \multicolumn{1}{l|}{$\hat{I}_4$} & \multicolumn{1}{l|}{$\hat{I}_4^{\ap}$} & \multicolumn{1}{l|}{$\hat{I}_4$ error} & \multicolumn{1}{l|}{$\hat{I}_6$} & \multicolumn{1}{l|}{$\hat{I}_6^{\ap}$} & \multicolumn{1}{l|}{$\hat{I}_6$ error} \\ \hline
\multicolumn{1}{|l|}{} & \multicolumn{1}{l|}{} & \multicolumn{1}{l|}{} & \multicolumn{1}{l|}{} & \multicolumn{1}{l|}{} & \multicolumn{1}{l|}{} & \multicolumn{1}{l|}{} \\ \hline
3 & 1.5 & 2.12597868844099 & 41.7\% & 32.4 & 7.00428372252487 & 78.4\% \\ \hline
4 & 1.77112299465241 & 2.11737368908595 & 19.5\% & 29.2857260386672 & 5.04095606861884 & 82.8\% \\ \hline
5 & 1.92727455514225 & 2.10358894768327 & 9.1\% & 24.3553395496129 & 3.85885436737017 & 84.2\% \\ \hline
6 & 1.97167595506035 & 2.09127008496943 & 6.1\% & 19.3570221116088 & 3.09958641182161 & 84.0\% \\ \hline
7 & 1.99917180579183 & 2.08108076102509 & 4.1\% & 15.8157978386411 & 2.57924397229046 & 83.7\% \\ \hline
8 & 2.01319334170647 & 2.07273170306951 & 3.0\% & 13.2137920826172 & 2.20346441160235 & 83.3\% \\ \hline
9 & 2.02020128573558 & 2.06584181515105 & 2.3\% & 11.2455089383337 & 1.92065635218756 & 82.9\% \\ \hline
10 & 2.02414460391682 & 2.06009120107809 & 1.8\% & 9.75173060605997 & 1.70073355331074 & 82.6\% \\ \hline
11 & 2.02625080279343 & 2.0552339688013 & 1.4\% & 8.58562698480205 & 1.52513914797188 & 82.2\% \\ \hline
12 & 2.02719976576841 & 2.05108476391051 & 1.2\% & 7.65247243254821 & 1.38187395446424 & 81.9\% \\ \hline
\multicolumn{1}{|l|}{} & \multicolumn{1}{l|}{} & \multicolumn{1}{l|}{} & \multicolumn{1}{l|}{} & \multicolumn{1}{l|}{} & \multicolumn{1}{l|}{} & \multicolumn{1}{l|}{} \\ \hline
\multicolumn{1}{|l|}{$\infty$} & 2 & \multicolumn{1}{l|}{} & \multicolumn{1}{l|}{} & 0 & \multicolumn{1}{l|}{} & \multicolumn{1}{l|}{} \\ \hline
\end{tabular}
\caption{Exact value of the moments and their approximation from the diagrammatic technique for $\Gamma=6$.}
\label{tab:G6Idiag}
\end{table}

\begin{table}[htbp]
\begin{tabular}{|r|l|r|l|l|r|l|}
\hline
$N$ & $\hat{I}_4$ & \multicolumn{1}{l|}{$\hat{I}_4^{\ap}$} & $\hat{I}_4$ error & $\hat{I}_6$ & \multicolumn{1}{l|}{$\hat{I}_6^{\ap}$} & $\hat{I}_6$ error \\ \hline
 &  & \multicolumn{1}{l|}{} &  &  & \multicolumn{1}{l|}{} &  \\ \hline
\multicolumn{1}{|r|}{3} & \multicolumn{1}{r|}{5.9504132231405} & 7.336835008 & \multicolumn{1}{r|}{23.3\%} & \multicolumn{1}{r|}{103.537190082645} & -127.1369698 & \multicolumn{1}{r|}{222.8\%} \\ \hline
\multicolumn{1}{|r|}{4} & \multicolumn{1}{r|}{5.35216744227873} & 6.1509483 & \multicolumn{1}{r|}{14.9\%} & \multicolumn{1}{r|}{57.4074000134889} & -71.80581066 & \multicolumn{1}{r|}{225.1\%} \\ \hline
\multicolumn{1}{|r|}{5} & \multicolumn{1}{r|}{5.31949584221798} & 5.585392260 & \multicolumn{1}{r|}{5.0\%} & \multicolumn{1}{r|}{30.083006000112} & -54.66045440 & \multicolumn{1}{r|}{281.7\%} \\ \hline
\multicolumn{1}{|r|}{6} & \multicolumn{1}{r|}{5.03608123946663} & 5.254867561 & \multicolumn{1}{r|}{4.3\%} & \multicolumn{1}{r|}{8.52435062302512} & -46.46113141 & \multicolumn{1}{r|}{645.0\%} \\ \hline
\multicolumn{1}{|r|}{7} & \multicolumn{1}{r|}{4.90673278031246} & 5.038212233 & \multicolumn{1}{r|}{2.7\%} & \multicolumn{1}{r|}{-1.73452473741117} & -41.68739389 & \multicolumn{1}{r|}{2303.4\%} \\ \hline
 8 &  & 4.885278337 &  &  & -38.57330645 &  \\ \hline
 9 &  & 4.771577915 &  &  & -36.38521278 &  \\ \hline
 10 &  & 4.683739174 &  &  & -34.76512324 &  \\ \hline
 11 &  & 4.613844211 &  &  & -33.51802809 &  \\ \hline
 12 &  & 4.556906725 &  &  & -32.52880274 &  \\ \hline
 13 &  & 4.509630618 &  &  & -31.72517539 &  \\ \hline
 14 &  & 4.469749993 &  &  & -31.05951906 &  \\ \hline
 15 &  & 4.435655939 &  &  & -30.49918829 &  \\ \hline
 16 &  & 4.406174568 &  &  & -30.02106637 &  \\ \hline
 17 &  & 4.380429327 &  &  & -29.60832983 &  \\ \hline
 18 &  & 4.357752484 &  &  & -29.24844597 &  \\ \hline
 19 &  & 4.337626449 &  &  & -28.93188851 &  \\ \hline
 20 &  & 4.319643825 &  &  & -28.65128805 &  \\ \hline
  &  & \multicolumn{1}{l|}{} &  &  & \multicolumn{1}{l|}{} &  \\ \hline
$\infty$ & \multicolumn{1}{r|}{4} & \multicolumn{1}{l|}{} &  & \multicolumn{1}{r|}{-24} & \multicolumn{1}{l|}{} &  \\ \hline
\end{tabular}
\caption{Exact value of the moments and their approximation from the diagrammatic technique for $\Gamma=8$.}
\label{tab:G8Idiag}
\end{table}

\begin{table}[htbp]
\begin{tabular}{|r|r|r|r|r|r|r|}
\hline
\multicolumn{1}{|l|}{$N$} & \multicolumn{1}{l|}{$\hat{I}_4$} & \multicolumn{1}{l|}{$\hat{I}_4^{\ap}$} & \multicolumn{1}{l|}{$\hat{I}_4$ error} & \multicolumn{1}{l|}{$\hat{I}_6$} & \multicolumn{1}{l|}{$\hat{I}_6^{\ap}$} & \multicolumn{1}{l|}{$\hat{I}_6$ error} \\ \hline
\multicolumn{1}{|l|}{} & 2 & \multicolumn{1}{l|}{} & \multicolumn{1}{l|}{} & 3 & \multicolumn{1}{l|}{} & \multicolumn{1}{l|}{} \\ \hline
\multicolumn{1}{|l|}{} & \multicolumn{1}{l|}{} & \multicolumn{1}{l|}{} & \multicolumn{1}{l|}{} & \multicolumn{1}{l|}{} & \multicolumn{1}{l|}{} & \multicolumn{1}{l|}{} \\ \hline
2 & -0.666666666666667 & -0.6317574181 & 5.236\% & -0.8 & -0.752415111 & 5.95\% \\ \hline
3 & -0.9 & -0.8700339821 & 3.330\% & -1.35 & -1.337880192 & 0.90\% \\ \hline
4 & -1.06666666666667 & -1.042131679 & 2.300\% & -1.82857142857143 & -1.864218645 & 1.95\% \\ \hline
5 & -1.19047619047619 & -1.170437198 & 1.683\% & -2.23214285714286 & -2.312761744 & 3.61\% \\ \hline
6 & -1.28571428571429 & -1.26919496 & 1.285\% & -2.57142857142857 & -2.689601017 & 4.60\% \\ \hline
7 & -1.36111111111111 & -1.347327155 & 1.013\% & -2.85833333333333 & -3.00618709 & 5.17\% \\ \hline
8 & -1.42222222222222 & -1.410579495 & 0.819\% & -3.1030303030303 & -3.27362061 & 5.50\% \\ \hline
9 & -1.47272727272727 & -1.462780494 & 0.675\% & -3.31363636363636 & -3.501250962 & 5.66\% \\ \hline
10 & -1.51515151515152 & -1.50656529 & 0.567\% & -3.4965034965035 & -3.69658924 & 5.72\% \\ \hline
11 & -1.55128205128205 & -1.543801073 & 0.482\% & -3.65659340659341 & -3.865574853 & 5.72\% \\ \hline
12 & -1.58241758241758 & -1.57584505 & 0.415\% & -3.7978021978022 & -4.012891553 & 5.66\% \\ \hline
13 & -1.60952380952381 & -1.603706117 & 0.361\% & -3.92321428571429 & -4.142245077 & 5.58\% \\ \hline
14 & -1.63333333333333 & -1.628149088 & 0.317\% & -4.03529411764706 & -4.256585403 & 5.48\% \\ \hline
15 & -1.65441176470588 & -1.649763932 & 0.281\% & -4.13602941176471 & -4.358278405 & 5.37\% \\ \hline
16 & -1.67320261437909 & -1.669012791 & 0.250\% & -4.22703818369453 & -4.449236580 & 5.26\% \\ \hline
17 & -1.69005847953216 & -1.686262682 & 0.225\% & -4.30964912280702 & -4.531018304 & 5.14\% \\ \hline
18 & -1.70526315789474 & -1.701808689 & 0.203\% & -4.38496240601504 & -4.604903343 & 5.02\% \\ \hline
19 & -1.71904761904762 & -1.715890707 & 0.184\% & -4.4538961038961 & -4.671950573 & 4.90\% \\ \hline
20 & -1.73160173160173 & -1.72870573 & 0.167\% & -4.51722190852626 & -4.733042400 & 4.78\% \\ \hline
21 & -1.74308300395257 & -1.740417020 & 0.153\% & -4.57559288537549 & -4.788919165 & 4.66\% \\ \hline
22 & -1.7536231884058 & -1.751160988 & 0.140\% & -4.6295652173913 & -4.840206004 & 4.55\% \\ \hline
23 & -1.76333333333333 & -1.761052508 & 0.129\% & -4.67961538461538 & -4.88743397 & 4.44\% \\ \hline
24 & -1.77230769230769 & -1.770188989 & 0.120\% & -4.72615384615385 & -4.931056780 & 4.34\% \\ \hline
25 & -1.78062678062678 & -1.778653568 & 0.111\% & -4.76953601953602 & -4.971464120 & 4.23\% \\ \hline
26 & -1.78835978835979 & -1.786517625 & 0.103\% & -4.81007115489874 & -5.008992411 & 4.14\% \\ \hline
27 & -1.79556650246305 & -1.793842793 & 0.096\% & -4.84802955665025 & -5.043933457 & 4.04\% \\ \hline
28 & -1.80229885057471 & -1.800682557 & 0.090\% & -4.88364849833148 & -5.076541501 & 3.95\% \\ \hline
29 & -1.80860215053763 & -1.807083561 & 0.084\% & -4.91713709677419 & -5.107039000 & 3.86\% \\ \hline
30 & -1.81451612903226 & -1.813086665 & 0.079\% & -4.94868035190616 & -5.135621381 & 3.78\% \\ \hline
31 & -1.82007575757576 & -1.818727815 & 0.074\% & -4.97844251336898 & -5.16246097 & 3.70\% \\ \hline
32 & -1.825311942959 & -1.824038756 & 0.070\% & -5.00656990068755 & -5.187710286 & 3.62\% \\ \hline
\multicolumn{1}{|l|}{} & \multicolumn{1}{l|}{} & \multicolumn{1}{l|}{} & \multicolumn{1}{l|}{} & \multicolumn{1}{l|}{} & \multicolumn{1}{l|}{} & \multicolumn{1}{l|}{} \\ \hline
\multicolumn{1}{|l|}{$\infty$} & -2 & \multicolumn{1}{l|}{} & \multicolumn{1}{l|}{} & -6 & \multicolumn{1}{l|}{} & \multicolumn{1}{l|}{} \\ \hline
\end{tabular}
\caption{Exact value of the moments and their approximation from the diagrammatic technique for $\Gamma=2$.}
\label{tab:G2Idiag}
\end{table}

%%%%%%%%%%%%%%%%%%%%%%%%%%%%%%%%%%%%%%%%%%%%%
\break  % to avoid tables from previous section overlapping in the
        % following section

\section{Fluctuation of linear statistics}\label{S5}
It was remarked in the Introduction that the sum rules for the moments of $\rho_{(2)}^T(\vec{r},\vec{0})$
are related to the average value of the linear statistic $\sum_{l=1}^N |\vec{r}_l |^{2n}$.
Here we point out that in the case of the soft disk geometry and $n=1$ the exact  distribution function for this
linear statistic is easily evaluated. Two features of the exact distribution are singled out for further study.
One relates to the O$(1)$ correction term to the mean. We are able to deduce the value of this correction
to the mean for a general smooth radial linear statistic. The other is that 
for large $N$ the exact
distribution tends to a Gaussian, with mean and variance as can be anticipated from a Coulomb gas
viewpoint \cite{Fo99}. We will show that a mechanism for such Gaussian fluctuations in the case of
general rotationally invariant linear statistics can be identified from the viewpoint of the expansions
(\ref{2.1a}) and (\ref{3.1a}).

First, with $\langle \cdot \rangle_{\rm plane}$ denoting an average with respect to the Boltzmann factor (\ref{15.110a}),
we observe that a simple scaling of the coordinates $\vec{r}_j$ gives
\begin{equation}\label{6.1}
\Big \langle e^{i k \sum_{j=1}^N |\vec{r}_j|^2} \Big \rangle_{\rm plane} =
\Big ( 1 - {2 i k \over \Gamma \pi \rho_b} \Big )^{-N - \Gamma N (N-1)/4}.
\end{equation}
Coulomb gas theory \cite{Fo99} tells us that setting $R=1$, so $\rho_b = N/\pi$, then taking $N \to \infty$, the distribution function
of a radial linear statistic $A = \sum_{l=1}^N g(r_l)$ should become the Gaussian
\begin{equation}\label{99}
\Big \langle e^{i k \sum_{j=1}^N g(r_l)}  \Big \rangle_{\rm plane} \mathop{\sim}\limits_{N \to \infty}
e^{i k \mmu_N} e^{- k^2 \sigma^2/2}
\end{equation}
with
\begin{align}\label{6.1a}
\mmu_N & = 2N \int_0^1 r g(r) \, dr + {\rm O}(1)  \nonumber \\
\sigma^2 & = {1 \over \Gamma} \int_0^1 r (g'(r))^2 \, dr + {\rm o}(1).
\end{align}

\subsection{Universal correction to the mean $\mmu_N$}
According to (\ref{6.1}), with $\rho_b = N/\pi$,
\begin{align}\label{E}
\Big \langle e^{i k \sum_{j=1}^N |\vec{r}_j|^2} \Big \rangle_{\rm plane} & =
\exp \bigg ( {i k N \over 2} + {2 i k \over \Gamma} \Big ( 1 - {\Gamma \over 4} \Big ) - {k^2  \over  2 \Gamma} +   {\rm O}\Big ( {1 \over N} \Big ) \bigg ).
\end{align}
This is of the form (\ref{99}) with
\begin{align}
\mmu_N & = {N \over 2} + {2  \over \Gamma} \Big (1 - {\Gamma \over 4} \Big )  \label{6.1b} \\
\sigma^2 & =  {1 \over  \Gamma} , \label{6.1c} 
\end{align}
and (\ref{6.1b}), (\ref{6.1c}) are in turn consistent with (\ref{6.1a}) in the case $g(r) = r^2$. The expression (\ref{6.1b}) is an identity for all $N$. But with
$g(r) = r^2$,
\begin{equation}
\mmu_N = \int_{\mathbb R^2} r^2 \rho_{(1)}(r) \, d^2 \vec{r}.  
\end{equation}
Writing $\tilde{\rho}_b = \rho_b \chi_{0 < r < R}$, where $\chi_A = 1$ if $A$ is true, and $\chi_A = 0$ otherwise, this can be written
\begin{equation}\label{3.30a}
\mmu_N = {\pi \over 2} R^4 \rho_b + \int_{\mathbb R^2} r^2 (\rho_{(1)}(\vec{r}) - \tilde{\rho}_b) \, d^2 \vec{r}.
\end{equation}
Recalling that in (\ref{6.1b}), $\rho_b = N/\pi$ and $R=1$, comparison of (\ref{3.30a}) and (\ref{6.1b}) gives the sum rule
\begin{equation}\label{con}
 \int_{\mathbb R^2} r^2 (\rho_{(1)}(\vec{r}) - \tilde{\rho}_b) \, d^2 \vec{r} = R^2 {2 \over \Gamma} \Big ( 1 - {\Gamma \over 4} \Big ).
 \end{equation}
 
 We remark that an alternative derivation of (\ref{con}), valid for $\Gamma$ even, 
is to use  (\ref{3.1}) and (\ref{3.1x}), together with 
the recurrence relation for the gamma function. 
We remark too that
(\ref{con}) is the soft disk version of the sum rule for the 2dOCP in a disk with hard wall boundary conditions \cite{CFG80},
 \begin{equation}
 \rho_{(1)}(0) - \Big (1 - {\Gamma \over 4} \Big ) \rho_b = - {\Gamma \rho_b^2 \pi^2 \over N}
 \int_0^R r^3 \Big ( \rho_{(1)}(R - r) - \rho_b \Big ) \, dr,
 \end{equation}
where $\rho_{(1)}(r)$ is measured inward from the boundary. This latter viewpoint allows the $N \to \infty$ limit of (\ref{con}) to be formulated. 
Thus on the LHS change to polar coordinates, and further change variables $r \mapsto R - r$.
Defining  $\rho_{(1)}^{\rm sw}(r) = \rho_{(1)}(R-r)$, corresponding to 
the density as measured inward from $r=R$ in the soft wall disk OCP, and noting that
charge neutrality requires
 \begin{equation}\label{cn}
\int^R_{-\infty} (R-r) (\rho_{(1)}^{\rm sw}(r) - \tilde{\rho_b} ) \, dr = 0
 \end{equation}
we see that
\begin{align}\label{hs}
\int_0^\infty r^3(\rho_{(1)}(r) - \tilde{\rho}_b) \, dr & =  \int^R_{-\infty} (R - r)^3(\rho_{(1)}^{\rm sw}(r) - \tilde{\rho}_b) \, dr
\nonumber \\
&= \int_{-\infty}^R (-2R^2 r + 3R r^2 - r^3)  (\rho_{(1)}^{\rm sw}(r) - \tilde{\rho_b} ) \, dr \nonumber \\
& =  -2 R^2 \int_{-\infty}^\infty r   (\rho_{(1)}^{\rm sw}(r) - \tilde{\rho_b} ) \, dr + {\rm O}(R)
\end{align}
(compare with working below (3.17) of \cite{TF99}).
Comparison with (\ref{con}) we thus have that in the thermodynamic limit
\begin{equation}\label{con1}
-2 \pi \Gamma  \int_{-\infty}^\infty r   (\rho_{(1)}^{\rm sw}(r) - \tilde{\rho_b} ) \, dr  =  \Big ( 1 - {\Gamma \over 4} \Big ).
\end{equation}
The sum rule (\ref{con1}) should be compared against the contact theorem \cite{CFG80} for the hard wall plasma,
\begin{equation}\label{con2}
\rho_{(1)}^{\rm hw}(0) -2 \pi \Gamma  \int_{0}^\infty r   (\rho_{(1)}^{\rm hw}(r) - \tilde{\rho_b} ) \, dr  =  \Big ( 1 - {\Gamma \over 4} \Big ),
\end{equation}
where $\rho_{(1)}^{\rm hw}(r)$ denotes the density as measured from the boundary.

The above working suggests that the first two terms in the asymptotic expansion of $\mmu_N$ in the case of $g(r) = r^m$, for $m \in \mathbb Z^+$,
admit a form analogous to that exhibited in (\ref{6.1b}). First, with this choice of $g(r)$, analogous to (\ref{3.30a}) we have
$$
\mmu_N = {2 \pi \over m+2} R^{m+2} \rho_b +2 \pi  \int_0^\infty
r^{m+1} (\rho_{(1)}(r) - \tilde{\rho}_b) \, dr
\,.
$$
Manipulation of the integral as in (\ref{hs}) now shows
$$
\mmu_N = {2 \pi \over m+2} R^{m+2} \rho_b - 2 \pi p R^m \int_{-\infty}^\infty r   (\rho_{(1)}^{\rm sw}(r) - \tilde{\rho_b} ) \, dr + {\rm O}(R^{m-1}\rho_b^{-1/2}).
$$
Substituting of (\ref{con1}) we then obtain
\begin{equation}\label{con3}
\mmu_N = {2 \pi \over m+2} R^{m+2} \rho_b + {m  \over \Gamma} \Big (1 - {\Gamma \over 4} \Big ) R^{m}  + {\rm O}(R^{m-1}\rho_b^{-1/2}).
\end{equation}
In particular, with $R =1$ and $\rho_b = N/\pi$, we have
\begin{equation}\label{con4}
\mmu_N = {2 N\over m+2}  + {m  \over \Gamma} \Big (1 - {\Gamma \over 4} \Big )   + {\rm O}\Big ( {1 \over N^{1/2} } \Big ).
\end{equation}
In this setting write $\rho_{(1)}(r) = (N/\pi)\chi_{0<r<1} + \kappa(r) + {\rm O}(N^{-1/2})$. Then it follows by comparing the definition of $\mmu_N$ in the case
 $g(r) = r^m$ to (\ref{con4}) that
 $$
 2 \pi \int_0^\infty r^{m+1} \kappa(r) \, dr =  {m  \over \Gamma} \Big (1 - {\Gamma \over 4} \Big ) .
 $$
 And it follows from this that $\kappa(r)$ is the distribution supported at $r=1$ given by
 \begin{equation}\label{kk}
 \kappa(r) =  {1  \over 2 \pi \Gamma} \Big (1 - {\Gamma \over 4} \Big ) {1 \over r} \delta'(r - 1)  .
\end{equation} 
Thus for $g(r)$ smooth in the neighbourhood of $r=1$, and with $R=1$ and $\rho_b = N/\pi$, we have that for large $N$
 \begin{equation}\label{kk1}
 \mmu_N = 2 \pi N \int_0^1 r g(r) \, dr + {1  \over  \Gamma} \Big (1 - {\Gamma \over 4} \Big ) g'(1) + {\rm O}(N^{-1/2}).
 \end{equation}

\subsection{Derivation using (\ref{3.1x})}
We will now consider the structure of
the Gaussian fluctuation formula (\ref{99}) itself.  It can readily be deduced at $\Gamma = 2$ as a consequence of the expansion formula
(\ref{3.1x}) for $I_{N,\Gamma}[g]$ \cite{Fo99}. Thus then $p=1$ and $c_\mu^{(N)}(1) = 1$ for $\mu = \delta_N$, and
$c_\mu^{(N)}(1) = 0$ otherwise. It is interesting to probe how (\ref{3.1x}) can lead to (\ref{99}) for general $\Gamma = 4p+2$.
We then have
\begin{equation}\label{2.11}
\Big \langle e^{i k \sum_{j=1}^N a(r_j) } \Big \rangle_{\rm plane} =
{ I_{N,\Gamma}[e^{i k a(r)}] \over I_{N,\Gamma}[1] } 
\end{equation}
where, with $G_{\mu_l}$ specified by (\ref{gg})
\begin{equation}\label{2.12}
 I_{N,\Gamma}[e^{i k a(r)}]  = N! \pi^N
 \sum_\mu \Big ( c_\mu^{(N)}(2p+1) \Big )^2
 \prod_{l=1}^N G_{\mu_l} [e^{- N \Gamma r^2/2} e^{i k a(r^2)}] ,
 \end{equation}
and according to (\ref{2.13a})
\begin{equation}\label{2.13b}
{\Gamma \over 2} (N - 1) \ge \mu_1 > \mu_2 > \cdots > \mu_N \ge 0
\end{equation}
with $|\mu | = \Gamma N (N - 1)/4$. Let's introduce the scaled variables $\tilde{\mu}_l := 2 \mu_l/N \Gamma$ so that to leading order in
$N$
$$
1 \ge \tilde{\mu}_1 > \tilde{\mu}_2 > \cdots > \tilde{\mu}_N \ge 0.
$$

In terms of these scaled variables, and upon the change of variables $r^2 = s$, we see
$$
 G_{\mu_l} [e^{- N \Gamma r^2/2} e^{i k a(r^2)}]  = \int_0^\infty e^{-(\Gamma N /2)(s - \tilde{\mu}_l \log s) + i k a (\sqrt{s})} \, ds.
 $$
 For large $N$ the maximum of the exponent occurs at $s = \tilde{\mu}_l$. Expanding the integrand about this point and completing the square 
 shows that for large $N$
 \begin{equation}
G_{\mu_l} [e^{- N \Gamma r^2/2} e^{i k a(r^2)}]  \sim
G_{\mu_l} [e^{- N \Gamma r^2/2} ] \exp \Big ( i k a(\sqrt{\tilde{\mu}_l}) -
{k^2 \over 4 \Gamma N} (a'(\sqrt{ \tilde{\mu}_l}) )^2 \Big ).   
 \end{equation}
To proceed further we hypothesize that the values of $\{ \tilde{\mu}_l \}$ for which the $c_\mu^{(N)}(2p +1)$ in (\ref{3.1x}) are
nonzero are
uniformly distributed (or at least this situation dominates). Then we have
\begin{eqnarray}
\lefteqn{ \prod_{l=1}^N  G_{\mu_l} [e^{- N \Gamma r^2/2} e^{i k
      a(r^2)}]  } 
\nonumber\\
&& \sim \prod_{l=1}^N  G_{\mu_l}   [e^{- N \Gamma r^2/2} ]
\exp \Big ( i k N \int_0^1 a(\sqrt{u}) \, du - {k^2 \over 4 \Gamma} \int_0^1
( a'(\sqrt{u}))^2 \, du \Big ).
\end{eqnarray}
Substituting in (\ref{2.12}) then substituting the result in
(\ref{2.11}) reclaims (\ref{99}).

\subsection{Test near $\Gamma=2$}
\label{SsubGamma2}

In this subsection we verify~(\ref{con4}) for $\Gamma=2$ and for $\Gamma$
close to 2, where explicit analytical calculations of $\mmu_N$ can be
done. When $\Gamma=2$, the density profile is~\cite[\S 15]{Fo10}
\begin{equation}
  \rho_{(1)}(r)=\rho_b e^{-\pi \rho_b r^2} \sum_{k=0}^{N-1}
  \frac{(\pi\rho_b r^2)^k}{k!}
=  \frac{N \Gamma(N,Nr^2)}{\pi(N-1)!}
\,,
\end{equation}
where $\pi\rho_b=N$.  Then, with $m=2n$,
\begin{equation}
  \mmu_N=\int_{\mathbb{R}^2} r^{2n} \rho_{(1)}(r) d^2\vec{r} 
  = \frac{1}{N^n}\sum_{k=0}^{N-1} \frac{(k+n)!}{k!} 
  = \frac{N(N+n)!}{N^n(1+n)N!}\,.
\end{equation}
Now, when $N\to\infty$,
\begin{equation}
  \left.\mmu_N\right|_{\Gamma=2}=\frac{N}{n+1}+\frac{n}{2}+\frac{n(n-1)(3n+2)}{N}+\OO(1/N^2)
\end{equation}
which satisfies~(\ref{con4}) when $\Gamma=2$. It is interesting to
notice that there are no $\OO(N^{-1/2})$ corrections to $\mmu_N$ at
$\Gamma=2$. This is, however, a special feature of the $\Gamma=2$
case, as for general $\Gamma$ there will be non zero $\OO(N^{-1/2})$
corrections (see sec.~\ref{S:moment-numerics}). A similar situation
happens for the finite-size corrections to the free energy of the
2dOCP in the soft disk at $\Gamma=2$~\cite{TF99}.

For $\Gamma$ close to 2, one can perform an expansion in powers of
$\Gamma-2$ of the density profile, to compute $\mmu_{N}$, similar to a
computation done in~\cite{Jan81}. To the linear order in $\Gamma-2$,
the density is given by
$\rho_{(1)}(\vec{r};\Gamma)=\rho_{(1)}(\vec{r};2)-\frac{\Gamma-2}{2}\langle
\hat\rho(\vec{r}) \beta H \rangle^T+\OO((\Gamma-2)^2)$, where the
truncated average is taken with the Boltzmann factor~(\ref{15.110a})
at $\Gamma=2$,  $H$ is the potential energy of the 2dOCP and
$\hat{\rho}(\vec{r})$ is the microscopic density. Explicitly,
\begin{eqnarray}
\langle
\hat\rho(\vec{r}) \beta H \rangle^T &=&
\iint_{\mathbb{R}^2\times\mathbb{R}^2} 
 \left[\rho_{(3)}(\vec{r}_1, \vec{r}_2,
\vec{r})-\rho_{(1)}(\vec{r})\rho_{(2)}(\vec{r}_1,\vec{r}_2)\right]
v(\vec{r}_1,\vec{r}_2) \,d^2\vec{r}_1 d^2\vec{r}_2
+\pi \rho_b r^2 \rho_{(1)}(\vec{r})
\nonumber\\
&+& 2\int_{\mathbb{R}^2} \rho_{(2)}(\vec{r}_1,\vec{r})
v(\vec{r}_1,\vec{r})\,d^2\vec{r}_1
+\pi \rho_b \int_{\mathbb{R}^2} r_1^2 \left[\rho_{(2)}(\vec{r}_1,
\vec{r})
-\rho_{(1)}(\vec{r}_1)\rho_{(1)}(\vec{r})\right] \,d^2\vec{r}_1
\nonumber\\
\label{eq:rhoH}
\end{eqnarray}
where the density and correlations in the RHS are evaluated at
$\Gamma=2$ and $v(\vec{r},\vec{r}\,')$ is the
Coulomb pair potential~(\ref{eq:Coulomb-plane}). At $\Gamma=2$, the
correlation functions have the simple structure~\cite{Jan81} \cite[\S 15.3]{Fo10} 
\begin{equation}
  \label{eq:correl-G2}
  \rho_{(\ell)}(\vec{r}_1,\vec{r}_2,\ldots,\vec{r}_{\ell})=\rho_b^{\ell}
  \det (K(z_{i},z_{j}))_{1\leq i,j\leq\ell}
\,,
\end{equation}
with
\begin{equation}
  K(z,z')=e^{-(|z|^2+|z'|^2)/2}\sum_{k=0}^{N-1} \frac{(z\bar{z}')^k}{k!}\,,
\end{equation}
and $z=\sqrt{\pi\rho_b} r e^{i\phi}$, where $(r,\phi)$ are the polar
coordinates of $\vec{r}$. To perform the integrals in~(\ref{eq:rhoH}),
it is convenient to expand the Coulomb
potential~(\ref{eq:Coulomb-plane}) in a Fourier series in the polar
angle
\begin{equation}
  \label{eq:Coulomb-Fourier}
  v(\vec{r}_1,\vec{r}_2)=\sum_{m=1}^{\infty} \frac{1}{m}
  \left(\frac{r_<}{r_>}\right)^m
  \cos[m(\phi_1-\phi_2)]-\log\frac{r_>}{L}
  \,,
\end{equation}
with $r_>=\max(r_1,r_2)$ and
$r_<=\min(r_1,r_2)$. Replacing~(\ref{eq:correl-G2})
and~(\ref{eq:Coulomb-Fourier}) in~(\ref{eq:rhoH}), after much
algebra, one obtains
\begin{eqnarray}
 \rho_{(1)}(\vec{r};\Gamma)&=&   \rho_{(1)}(\vec{r};2) 
  \nonumber\\
 &\hspace{-10mm}-& \hspace{-8mm}(\Gamma-2) \rho_b e^{-|z|^2}\Bigg\{
   \sum_{k_1=0}^{N-1} \sum_{k_2=0\atop k_2\neq k_1}^{N-1}
   \frac{|z|^{2k_2}}{2k_1!(k_2!)^2} \J(k_1,k_2)
   +\sum_{k_1=0}^{N-1}
   \sum_{k_2=k_1+1}^{N-1}\frac{\I(k_1,k_2)}{k_1!k_2!(k_2-k_1)}
   \left[\frac{|z|^{2k_2}}{k_2!}+\frac{|z|^{2k_1}}{k_1!}\right]
   \nonumber\\
   &-& \sum_{k_1=0}^{N-1}\sum_{k_2=k_1+1}^{N-1}
   \frac{|z|^{2k_1}\gamma(k_2+1,|z|^2)
     +|z|^{2k_2}\Gamma(k_1+1,|z|^2)}{(k_2-k_1)k_1!k_2!}
   \nonumber\\
   &-&\sum_{k_2=0}^{N-1} \frac{|z|^{2k_2}}{2\,k_2!}
   \sum_{k_1=0}^{N-1} \frac{\gamma(k_1,|z|^2)\log (|z|^2)+ \int_{|z|^2}^{\infty} e^{-t}
     t^{k_1} \log t \,dt }{k_1!}
   -\sum_{k_1=0}^{N-1} \frac{|z|^{2k_1}}{2\,k_1!}\left[k_1+1-|z|^2\right]
   \nonumber\\
   &+&\sum_{k_1=0}^{N-1} \frac{|z|^{2k_1}}{2(k_1!)^2}
   \left[ \gamma(k_1+1,|z|^2)\log (|z|^2) + \int_{|z|^2}^{\infty} e^{-t}
     t^{k_1} \log t \,dt \right]\Bigg\}+\OO((\Gamma-2)^2)
   \,,
\end{eqnarray}
where we have defined
\begin{equation}
  \I(k_1,k_2)=\iint_{0\leq t_2 < t_1} e^{-t_1-t_2} t_1^{k_1} t_2^{k_2}
  \,dt_1dt_2
  \,,
\end{equation}
and
\begin{equation}
  \J(k_1,k_2)=\int_{0}^{\infty}\int_{0}^{\infty}
  e^{-t_1-t_2} t_1^{k_1} t_2^{k_2} \log(\max(t_1 ,t_2)) \,dt_1dt_2
  \,.
\end{equation}
Some useful properties, as well as the asymptotic expansion when
$k_1\to\infty$ and $k_2\to\infty$ of $\I(k_1,k_2)$ and $\J(k_1,k_2)$,
are discussed in the appendix.

Consequently, with $\pi\rho_b=N$, the $2n$-moment of the density profile is
\begin{eqnarray}
  \mmu_N&=&  \left.\mmu_N\right|_{\Gamma=2} 
  -(\Gamma-2) N^{-n}\Bigg\{
  \frac{n}{2}\sum_{k_1=0}^{N-1}\frac{(k_1+n)!}{k_1!}
\nonumber\\
&+&
\sum_{k_1=0}^{N-1}
   \sum_{k_2=k_1+1}^{N-1}\frac{
   \frac{(k_1+n)!}{k_1!}\I(k_1,k_2)-\I(k_1+n,k_2)
     +\frac{(k_2+n)!}{k_2!}\I(k_1,k_2)-\I(k_1,k_2+n)
     }{k_1!k_2!(k_2-k_1)}\Bigg\}
\nonumber\\
&+&\sum_{k_1=0}^{N-1} \sum_{k_2=0\atop k_2\neq k_1}^{N-1}
\frac{1}{2\,k_1!k_2!}
\left[\frac{(k_1+n)!}{k_1!}\J(k_1,k_2)-\J(k_1+n,k_2)\right]
+\OO((\Gamma-2)^2)\,.
\label{eq:M1}
\end{eqnarray}
In the last term of~(\ref{eq:M1}) it is convenient to use the
recurrence relation~(\ref{eq:Jrec}) and write
\begin{eqnarray}
  \frac{(k_1+n)!}{k_1!}\J(k_1,k_2)-\J(k_1+n,k_2)&=&-\sum_{\ell=0}^{n-1}
\frac{(k_1+n)!}{k_1!(k_1+1+\ell)}
\I(k_1,k_2)
\nonumber\\
&-&\sum_{\ell=0}^{n-1}
\frac{(k_1+n)!}{(k_1+1+\ell)!}
\left[\I(k_1+\ell,k_2)
-\frac{(k_1+\ell)!}{k_1!}\I(k_1,k_2)\right]
\,.
\nonumber\\
\end{eqnarray}
Consequently $\mmu_N$ is given by
$\mmu_N=\left.\mmu_N\right|_{\Gamma=2} -(\Gamma-2)(\tilde{m}_1+\tilde{m}_2+\tilde{m}_3)
+\OO((\Gamma-2)^2)$
with
\begin{equation}
  \label{eq:m1}
  \tilde{m}_1=\frac{N^{-n}}{2}\left[\sum_{k_1=0}^{N-1}n\frac{(k_1+n)!}{k_1!}
    - \sum_{k_1=0}^{N-1}
   \sum_{k_2=0 \atop k_2\neq k_1}^{N-1}
   \frac{1}{k_1!k_2!}\sum_{\ell=0}^{n-1}
\frac{(k_1+n)!}{k_1!(k_1+1+\ell)}
\I(k_1,k_2)
\right]
\,,
\end{equation}
\begin{equation}
  \label{eq:m2}
  \tilde{m}_2=- \frac{N^{-n}}{2}\sum_{k_1=0}^{N-1}
   \sum_{k_2=0 \atop k_2\neq k_1}^{N-1}
   \frac{1}{k_1!k_2!}
\sum_{\ell=0}^{n-1}
\frac{(k_1+n)!}{(k_1+1+\ell)!}
\left[\I(k_1+\ell,k_2)
-\frac{(k_1+\ell)!}{k_1!}\I(k_1,k_2)\right]
\,,
\end{equation}
and
\begin{equation}
  \label{eq:m3}
  \tilde{m}_3=N^{-n}
  \sum_{k_1=0}^{N-1}
   \sum_{k_2=k_1+1}^{N-1}\frac{
   \frac{(k_1+n)!}{k_1!}\I(k_1,k_2)-\I(k_1+n,k_2)
     +\frac{(k_2+n)!}{k_2!}\I(k_1,k_2)-\I(k_1,k_2+n)
     }{k_1!k_2!(k_2-k_1)}
   \,.
\end{equation}
In~(\ref{eq:m1}) it is convenient to split the sum on $k_1$ and $k_2$
into two sums for $k_1< k_2$ and $k_1>k_2$ and in the latter
use~(\ref{eq:Isymmetry}) to find
$\tilde{m}_1=\tilde{m}_1^{a}+\tilde{m}_1^{b}$ with
\begin{equation}
  \label{eq:m1a}
  \tilde{m}_1^{a}=\frac{N^{-n}}{2}\sum_{k_1=0}^{N-1}
    \left(\frac{(k_1+n)!\,n}{k_1!}-k_1\sum_{\ell=0}^{n-1}
      \frac{(k_1+n)!}{k_1!(k_1+1+\ell)}\right)
\,,
\end{equation}
and
\begin{equation}
 \label{eq:m1b}
\tilde{m}_1^{b}= -\frac{N^{-n}}{2} \sum_{k_1=0}^{N-1}
   \sum_{k_2=k_1+1}^{N-1}
   \frac{\I(k_1,k_2)}{k_1!k_2!}
\sum_{\ell=0}^{n-1}
\left(
\frac{(k_1+n)!}{k_1!(k_1+1+\ell)}
-\frac{(k_2+n)!}{k_2!(k_2+1+\ell)}
\right)
\,.
\end{equation}
In~(\ref{eq:m1a}), the expression inside the sum is a polynomial of
$k_1$ of order $n-1$, therefore to obtain the order $\OO(N^{0})$ when
$N\to\infty$ of $\tilde{m}_1^a$ it is sufficient to consider the
leading order of the polynomial and replace the sum over $k_1$ by an integral,
\begin{eqnarray}
  \tilde{m}_1^a&=& \frac{N^{-n}}{2}
  \int_0^{N} dk_1 \sum_{\ell=0}^{n-1}\left[ (\ell+1) k_1^{n-1}
    +\OO(k_1^{n-2})\right]
  \nonumber\\
  &=& \frac{n+1}{4} + \OO(1/N)\,.
\end{eqnarray}

In the appendix it is shown that, for large $k_1$ and $k_2$,
\begin{equation}
  \label{eq:Iasympt2}
   \frac{\I(k_1,k_2)}{ k_1! k_2! }  \sim 
   \frac{1}{2}\,\erfc\left(\frac{k_2-k_1}{\sqrt{2(k_1+k_2)}}\right)
   \,.
\end{equation}
To obtain the leading order contribution in $N$ of
$\tilde{m}_1^b$ one should replace the sums over $k_1$ and
$k_2$ in~(\ref{eq:m1b}) by integrals, and realise
from~(\ref{eq:Iasympt2}) that the leading contribution will be
obtained for $k_1$ and $k_2$ laying in a strip following the line
$k_1=k_2$, with $k_1<k_2$, and of width of order $\sqrt{N}$. With this in mind, it is
useful to do a change of variables in the integrals, from $k_1$ and
$k_2$ to $k_{+}=k_1+k_2$, which goes along the strip, and
$u=(k_2-k_1)\sqrt{2k_+}$, which goes perpendicular to the strip. The
integral over $u$ will give $\OO(1)$ contributions, due to the fast
decay of the $\erfc$ function, while the one over $k_{+}$ will give a
contribution of order $\OO(N^{n})$. To obtain this, it is sufficient to
keep the leading order in $k_{+}$ in the integrand. In~(\ref{eq:m1b}),
the expression in the sum over $\ell$ is a polynomial of $k_1$ and
$k_2$ that vanishes when $k_1=k_2$, therefore it can be written as
\begin{equation}
  \sum_{\ell=0}^{n-1}
\left(
\frac{(k_1+n)!}{k_1!(k_1+1+\ell)}
-\frac{(k_2+n)!}{k_2!(k_2+1+\ell)}
\right)=-(k_2-k_1)Q(k_1,k_2)
\,,
\end{equation}
with $Q(k_1,k_2)$ a polynomial in $k_1$ and $k_2$ of order
$n-2$. Actually, to the leading order,
\begin{eqnarray}
  Q(k_1,k_2)&=&n\sum_{r=0}^{n-2} k_1^{n-2-r}k_2^{r} + \text{polynomial in
    $k_1$ and $k_2$ of order $n-3$}
\nonumber\\
&=&n(n-1)\left(\frac{k_{+}}{2}\right)^{n-2} + \OO(k_{+}^{n-3})
\,.
\end{eqnarray}
Then, replacing in~(\ref{eq:m1b}),
\begin{eqnarray}
  \tilde{m}_1^{b}&=& -\frac{N^{-n}}{2} \sum_{k_1=0}^{N-1}
   \sum_{k_2=k_1+1}^{N-1}
   \frac{\I(k_1,k_2)}{k_1!k_2!} (k_2-k_1) Q(k_1,k_2)
\nonumber\\
&=& \frac{N^{-n}}{4}
\int_{0}^{\infty} du \int_{0}^{2N} dk_{+}
u\, \erfc(u) \left[n(n-1) 2^{-(n-2)}k_{+}^{n-1} +\OO(k_{+}^{n-2})\right]+\OO(1/\sqrt{N})
\nonumber\\
&=& \frac{n-1}{4}+\OO(1/\sqrt{N})\,,
\end{eqnarray}
where the integral $\int_{0}^{\infty} u\, \erfc(u) \,du=1/4$ was
used. Putting together the last results, we find
\begin{equation}
  \tilde{m}_1=\frac{n}{2} + \OO(1/\sqrt{N})
  \,.
\end{equation}

Regarding the other contributions to $\mmu_N$, it can be noticed using
the recurrence relation~(\ref{eq:Irec}) that
$\tilde{m}_2+\tilde{m}_3=0$ when $n=0,1,2,$ and $3$. For $n>3$, the
leading order of these terms can be obtained following similar steps as the
ones done for $\tilde{m}_1$. The sums over $k_1$ and $k_2$ are
replaced by integrals over the variables $k_{+}$ and $u$ defined
above, and only the leading order in $k_{+}$ is
kept. From~(\ref{eq:Iasympt}), one can obtain that, for $k_{+}$ large,
\begin{equation}
  \frac{\I(k_1+\ell,k_2)-\frac{(k_1+\ell)!}{k_1!}\I(k_1,k_2)}{k_2!
    (k_1+\ell)!}
  = \frac{\ell e^{-u^2}}{\sqrt{2\pi k_{+}}}
  +\OO(1/k_{+})
  \,.
\end{equation}
Substituting into~(\ref{eq:m2}) shows
\begin{eqnarray}
  \tilde{m}_2&=& -{N^{-n}}{2}\int_{-\infty}^{\infty}  du
  \int_{0}^{2N} \frac{dk_{+}\sqrt{2k_{+}}}{2} \sum_{l=0}^{n-1}\frac{l
    e^{-u^2}}{\sqrt{2\pi k_{+}}} \left(\frac{k_{+}}{2}\right)^{n-1}
  +\OO(1/\sqrt{N})
  \nonumber\\
  &=& -\frac{n-1}{4}  +\OO(1/\sqrt{N})
\end{eqnarray}
For $\tilde{m}_3$, use
\begin{equation}
\frac{\frac{(k_1+n)!}{k_1!}\I(k_1,k_2)-\I(k_1+n,k_2)}{k_1!k_2!}
    =\frac{k_{+}^nne^{-u^2}}{2^n\sqrt{2\pi
        k_{+}}}\left[-1+\frac{(n-1)u}{\sqrt{2k_{+}}}+\OO(1/k_{+})\right]
\,,
\end{equation}
and
\begin{equation}
\frac{\frac{(k_2+n)!}{k_2!}\I(k_1,k_2)-\I(k_1,k_2+n)}{k_1!k_2!}
    =\frac{k_{+}^{n}ne^{-u^2}}{2^n\sqrt{2\pi k_{+}}}\left[1+\frac{(n-1)u}{\sqrt{2k_{+}}}+\OO(1/k_{+})\right]
\,,
\end{equation}
to obtain
\begin{eqnarray}
  \tilde{m}_3&=&N^{-n} \int_{0}^{\infty} du\int_{0}^{2N}
  \frac{dk_{+}\sqrt{2k_{+}}}{2}
\left(\frac{k_+}{2}\right)^{n}
\frac{2e^{-u^2} n(n-1) u}{u \sqrt{2\pi k_{+}} 2k_{+}}
+\OO(1/\sqrt{N})
\nonumber\\
&=&\frac{n-1}{4}+\OO(1/\sqrt{N})
\,.
\end{eqnarray}
Therefore $\tilde{m}_2+\tilde{m}_3=\OO(1/\sqrt{N})$ do not contribute to
order $\OO(1)$ in $\mmu_N$. Summing up all results,
\begin{equation}
  \mmu_{N}=\frac{N}{n+1}+\frac{n}{2}
  -(\Gamma-2)\frac{n}{2}+\OO(1/\sqrt{N})+\OO((\Gamma-2)^2)\,.
\end{equation}
This result is in agreement with the general formula~(\ref{con4}) when
it is expanded around $\Gamma=2$ to the linear order in $\Gamma-2$
with $m=2n$.
 
\subsection{Numerical results for $\Gamma=4$, $6$, and $8$}
\label{S:moment-numerics}

For $\Gamma$ even, by application of~(\ref{3.1}) or~(\ref{3.1x}), the
$2n$-moment of the density can be expressed as
\begin{equation}
  \mmu_{N}=\frac{(N\Gamma/2)^{-n}}{Z_{\soft}}
  \sum_{\mu} \frac{(c_{\mu}^{(N)}(\Gamma/2))^2}{\prod_i m_i!}
  \prod_{\ell=1}^{N}\mu_{\ell}!\sum_{k=1}^{N}
  \frac{(\mu_{k}+n)!}{\mu_{k}!}
\,,
\end{equation}
with
\begin{equation}
  Z_{\soft}=  \sum_{\mu} \frac{(c_{\mu}^{(N)}(\Gamma/2))^2}{\prod_i m_i!}
  \prod_{\ell=1}^{N}\mu_{\ell}!
\,,
\end{equation}
which is the partition function of the 2dOCP in the soft disk, up to a
multiplicative constant.  Tables \ref{mmuN-G4-n2}--\ref{mmuN-G8-n4}
show the numerical evaluation of $\mmu_N$ as well as a fit to
$\mmu_N=aN + b +c N^{-1/2} +d/N$, for $\Gamma=4$, $6$, and $8$, and
$n=2$, $3$, and $4$. These numerical results, and the proposed fit,
show that~(\ref{con4}) is indeed satisfied. The convergence is very
good for all moments for $\Gamma=4$ and $6$. For $\Gamma=8$,
unfortunately we were not able to perform the calculations beyond
$N=7$ particles, the results presented have not yet converged to the
$N\to\infty$ expected value. Notice that, contrary to the case
$\Gamma=2$, there are non-zero $\OO(N^{-1/2})$ finite-size corrections to
$\mmu_{N}$.

%%%%%%%%%%%%%%%%%%%%%%%%%%%%%%%%%%%%%%%%
\begin{table}[htbp]
\begin{center}
\begin{tabular}{|r|r|r|r|r|r|}
\hline
\multicolumn{1}{|l|}{$N$} & \multicolumn{1}{c}{$\mmu_N\qquad =$} &
\multicolumn{1}{c}{$\quad aN \quad +$} & \multicolumn{1}{c}{$\quad b
  \quad +$} & \multicolumn{1}{c}{$\quad c/\sqrt{N}\quad +$} & \multicolumn{1}{c|}{$d/N$} \\ \hline
\multicolumn{1}{|l|}{} & \multicolumn{1}{l|}{} & \multicolumn{1}{c|}{$a$} & \multicolumn{1}{c|}{$b$} & \multicolumn{1}{c|}{$c$} & \multicolumn{1}{c|}{$d$} \\ \hline
2 & 0.8125 & \multicolumn{1}{l|}{} & \multicolumn{1}{l|}{} & \multicolumn{1}{l|}{} & \multicolumn{1}{l|}{} \\ \hline
3 & 1.126262 & \multicolumn{1}{l|}{} & \multicolumn{1}{l|}{} & \multicolumn{1}{l|}{} & \multicolumn{1}{l|}{} \\ \hline
4 & 1.44563743218807 & \multicolumn{1}{l|}{} & \multicolumn{1}{l|}{} & \multicolumn{1}{l|}{} & \multicolumn{1}{l|}{} \\ \hline
5 & 1.76891109591098 & 0.336437 & -0.0742714 & 0.458953 & -0.221263 \\ \hline
6 & 2.09454890418255 & 0.333419 & -0.00242266 & 0.269645 & -0.0817588 \\ \hline
7 & 2.42171814295119 & 0.332974 & 0.0108746 & 0.230301 & -0.0491322 \\ \hline
8 & 2.74996856529295 & 0.33338 & -0.003688474 & 0.277614 & -0.0922608 \\ \hline
9 & 3.07901955876735 & 0.333458 & -0.00698637 & 0.289203 & -0.103695 \\ \hline
10 & 3.40868118671838 & 0.333396 & -0.00397741 & 0.277889 & -0.0917455 \\ \hline
11 & 3.73882025776555 & 0.333353 & -0.00169446 & 0.268779 & -0.0815306 \\ \hline
12 & 4.06934089384864 & 0.333341 & -0.000983289 & 0.265786 & -0.0779911 \\ \hline
13 & 4.4001722794716 & 0.333341 & -0.000982974 & 0.265784 & -0.0779894 \\ \hline
14 & 4.73126081887937 & 0.333342 & -0.00105077 & 0.266097 & -0.0783948 \\ \hline
\multicolumn{1}{|l|}{} & \multicolumn{1}{l|}{} & \multicolumn{1}{l|}{} & \multicolumn{1}{l|}{} & \multicolumn{1}{l|}{} & \multicolumn{1}{l|}{} \\ \hline
$\infty$ & \multicolumn{1}{l|}{} & $1/3$ & 0 & \multicolumn{1}{l|}{} & \multicolumn{1}{l|}{} \\ \hline
\end{tabular}
\end{center}
\caption{Fourth moment ($n=2$) of the density when $\Gamma=4$.}
\label{mmuN-G4-n2}
\end{table}

%%%%%%%%%%%%%%%%%%%%%%%%%%%%%%

\begin{table}[htbp]
\begin{center}
\begin{tabular}{|r|r|r|r|r|r|}
\hline
\multicolumn{1}{|c|}{$N$} & \multicolumn{1}{c}{$\mmu_N\qquad =$} & \multicolumn{1}{c}{$\quad aN \quad +$} & \multicolumn{1}{c}{$\quad b \quad +$} & \multicolumn{1}{c}{$\quad c/\sqrt{N} \quad +$} & \multicolumn{1}{c|}{$d/N$} \\ \hline
\multicolumn{1}{|l|}{} & \multicolumn{1}{l|}{} & \multicolumn{1}{c|}{$a$} & \multicolumn{1}{c|}{$b$} & \multicolumn{1}{c|}{$c$} & \multicolumn{1}{c|}{$d$} \\ \hline
2 & 0.890625 & \multicolumn{1}{l|}{} & \multicolumn{1}{l|}{} & \multicolumn{1}{l|}{} & \multicolumn{1}{l|}{} \\ \hline
3 & 1.08333333333333 & \multicolumn{1}{l|}{} & \multicolumn{1}{l|}{} & \multicolumn{1}{l|}{} & \multicolumn{1}{l|}{} \\ \hline
4 & 1.29792043399638 & \multicolumn{1}{l|}{} & \multicolumn{1}{l|}{} & \multicolumn{1}{l|}{} & \multicolumn{1}{l|}{} \\ \hline
5 & 1.52318930281443 & 0.249835 & 0.0492187 & 0.53598 & -0.0745192 \\ \hline
6 & 1.75432075208484 & 0.246695 & 0.123999 & 0.338947 & 0.0706775 \\ \hline
7 & 1.98916255005449 & 0.248062 & 0.0831756 & 0.459736 & -0.0294887 \\ \hline
8 & 2.22660014721048 & 0.249434 & 0.0339754 & 0.619579 & -0.175195 \\ \hline
9 & 2.46595156197257 & 0.249738 & 0.021223 & 0.664392 & -0.219408 \\ \hline
10 & 2.70676308425724 & 0.249748 & 0.0207354 & 0.666226 & -0.221345 \\ \hline
11 & 2.94871984780432 & 0.249774 & 0.0193389 & 0.671799 & -0.227593 \\ \hline
12 & 3.19159601492947 & 0.249824 & 0.0163325 & 0.684451 & -0.242556 \\ \hline
13 & 3.43522469549061 & 0.249871 & 0.0132299 & 0.69815 & -0.259556 \\ \hline
14 & 3.67947926593368 & 0.249905 & 0.0108374 & 0.709186 & -0.273865 \\ \hline
\multicolumn{1}{|l|}{} & \multicolumn{1}{l|}{} & \multicolumn{1}{l|}{} & \multicolumn{1}{l|}{} & \multicolumn{1}{l|}{} & \multicolumn{1}{l|}{} \\ \hline
$\infty$ & \multicolumn{1}{l|}{} & $0.25$ & 0 & \multicolumn{1}{l|}{} & \multicolumn{1}{l|}{} \\ \hline
\end{tabular}
\end{center}
\caption{Sixth moment ($n=3$) of the density when $\Gamma=4$.}
\label{mmuN-G4-n3}
\end{table}

%%%%%%%%%%%%%%%%%%%%%%%%%%%%%%%%%%%

\begin{table}[htbp]
\begin{center}
\begin{tabular}{|c|r|r|r|r|r|}
\hline
$N$ & \multicolumn{1}{c}{$\mmu_N\qquad =$} & \multicolumn{1}{c}{$\quad aN \quad +$} & \multicolumn{1}{c}{$\quad b \quad +$} & \multicolumn{1}{c}{$\quad c/\sqrt{N} \quad +$} & \multicolumn{1}{c|}{$d/N$} \\ \hline
 & \multicolumn{1}{l|}{} & \multicolumn{1}{c|}{$a$} & \multicolumn{1}{c|}{$b$} & \multicolumn{1}{c|}{$c$} & \multicolumn{1}{c|}{$d$} \\ \hline
2 & 1.21875 & \multicolumn{1}{l|}{} & \multicolumn{1}{l|}{} & \multicolumn{1}{l|}{} & \multicolumn{1}{l|}{} \\ \hline
3 & 1.25420875420875 & \multicolumn{1}{l|}{} & \multicolumn{1}{l|}{} & \multicolumn{1}{l|}{} & \multicolumn{1}{l|}{} \\ \hline
4 & 1.36950440777577 & \multicolumn{1}{l|}{} & \multicolumn{1}{l|}{} & \multicolumn{1}{l|}{} & \multicolumn{1}{l|}{} \\ \hline
5 & 1.51576558282311 & 0.188419 & 0.468569 & -0.269131 & 1.12729 \\ \hline
6 & 1.67718058035561 & 0.189303 & 0.447536 & -0.213712 & 1.08645 \\ \hline
7 & 1.84742557965232 & 0.194189 & 0.301703 & 0.217776 & 0.728635 \\ \hline
8 & 2.02343313425351 & 0.197002 & 0.20078 & 0.545661 & 0.42975 \\ \hline
9 & 2.20346793019457 & 0.197864 & 0.164668 & 0.672559 & 0.30455 \\ \hline
10 & 2.38645657506403 & 0.198269 & 0.145249 & 0.745579 & 0.227431 \\ \hline
11 & 2.57169539517539 & 0.198622 & 0.126216 & 0.821535 & 0.142267 \\ \hline
12 & 2.75870117826714 & 0.198925 & 0.108069 & 0.897903 & 0.0519525 \\ \hline
13 & 2.9471288020509 & 0.199158 & 0.0926932 & 0.965792 & -0.0322945 \\ \hline
14 & 3.13672349360904 & 0.199328 & 0.0804934 & 1.02207 & -0.10526 \\ \hline
 & \multicolumn{1}{l|}{} & \multicolumn{1}{l|}{} & \multicolumn{1}{l|}{} & \multicolumn{1}{l|}{} & \multicolumn{1}{l|}{} \\ \hline
$\infty$ & \multicolumn{1}{l|}{} & $0.2$ & 0 & \multicolumn{1}{l|}{} & \multicolumn{1}{l|}{} \\ \hline
\end{tabular}
\end{center}
\caption{Eighth moment ($n=4$) of the density when $\Gamma=4$.}
\label{mmuN-G4-n4}
\end{table}

%%%%%%%%%%%%%%%%%%%%%%%%%%%%%%%%%%%%%%%%%%%%%

\begin{table}[htbp]
\begin{center}
\begin{tabular}{|r|r|r|r|r|r|}
\hline
\multicolumn{1}{|c|}{$N$} & \multicolumn{1}{c}{$\mmu_N\qquad =$} & \multicolumn{1}{c}{$\quad aN \quad +$} & \multicolumn{1}{c}{$\quad b \quad +$} & \multicolumn{1}{c}{$\quad c/\sqrt{N} \quad +$} & \multicolumn{1}{c|}{$d/N$} \\ \hline
\multicolumn{1}{|l|}{} & \multicolumn{1}{l|}{} & \multicolumn{1}{c|}{$a$} & \multicolumn{1}{c|}{$b$} & \multicolumn{1}{c|}{$c$} & \multicolumn{1}{c|}{$d$} \\ \hline
3 & 0.829151732377539 & \multicolumn{1}{l|}{} & \multicolumn{1}{l|}{} & \multicolumn{1}{l|}{} & \multicolumn{1}{l|}{} \\ \hline
4 & 1.13999055712937 & \multicolumn{1}{l|}{} & \multicolumn{1}{l|}{} & \multicolumn{1}{l|}{} & \multicolumn{1}{l|}{} \\ \hline
5 & 1.45889183119874 & \multicolumn{1}{l|}{} & \multicolumn{1}{l|}{} & \multicolumn{1}{l|}{} & \multicolumn{1}{l|}{} \\ \hline
6 & 1.78179400313294 & 0.330681 & -0.23698 & -0.0197034 & 0.256393 \\ \hline
7 & 2.10668148567864 & 0.326556 & -0.113869 & -0.383961 & 0.55846 \\ \hline
8 & 2.43308749295152 & 0.335184 & -0.423364 & 0.621536 & -0.358107 \\ \hline
9 & 2.76069703430536 & 0.335669 & -0.443708 & 0.693027 & -0.428642 \\ \hline
10 & 3.08920070504297 & 0.333408 & -0.335393 & 0.285749 & 0.00150004 \\ \hline
11 & 3.41837572685644 & 0.33287 & -0.306358 & 0.169876 & 0.131419 \\ \hline
12 & 3.74807370935471 & 0.333069 & -0.318295 & 0.220112 & 0.0720079 \\ \hline
\multicolumn{1}{|l|}{} & \multicolumn{1}{l|}{} & \multicolumn{1}{l|}{} & \multicolumn{1}{l|}{} & \multicolumn{1}{l|}{} & \multicolumn{1}{l|}{} \\ \hline
\multicolumn{1}{|c|}{$\infty$} & \multicolumn{1}{l|}{} & $1/3$ & $-1/3$ & \multicolumn{1}{l|}{} & \multicolumn{1}{l|}{} \\ \hline
\end{tabular}
\end{center}
\caption{Fourth moment ($n=2$) of the density when $\Gamma=6$.}
\label{mmuN-G6-n2}
\end{table}

%%%%%%%%%%%%%%%%%%%%%%%%%%%%%%%%%%%%%%%%%%%%%

\begin{table}[htbp]
\begin{center}
\begin{tabular}{|r|r|r|r|r|r|}
\hline
\multicolumn{1}{|c|}{$N$} & \multicolumn{1}{c}{$\mmu_N\qquad =$} & \multicolumn{1}{c}{$\quad aN \quad +$} & \multicolumn{1}{c}{$\quad b \quad +$} & \multicolumn{1}{c}{$\quad c/\sqrt{N} \quad +$} & \multicolumn{1}{c|}{$d/N$} \\ \hline
\multicolumn{1}{|l|}{} & \multicolumn{1}{l|}{} & \multicolumn{1}{c|}{$a$} & \multicolumn{1}{c|}{$b$} & \multicolumn{1}{c|}{$c$} & \multicolumn{1}{c|}{$d$} \\ \hline
3 & 0.63878932696137 & \multicolumn{1}{l|}{} & \multicolumn{1}{l|}{} & \multicolumn{1}{l|}{} & \multicolumn{1}{l|}{} \\ \hline
4 & 0.852317437834435 & \multicolumn{1}{l|}{} & \multicolumn{1}{l|}{} & \multicolumn{1}{l|}{} & \multicolumn{1}{l|}{} \\ \hline
5 & 1.07602482170551 & \multicolumn{1}{l|}{} & \multicolumn{1}{l|}{} & \multicolumn{1}{l|}{} & \multicolumn{1}{l|}{} \\ \hline
6 & 1.30481948267947 & 0.238952 & -0.17935 & -0.0015329 & 0.306508 \\ \hline
7 & 1.5366992098612 & 0.241542 & -0.256653 & 0.227188 & 0.116838 \\ \hline
8 & 1.77130926929194 & 0.254997 & -0.739333 & 1.79534 & -1.31262 \\ \hline
9 & 2.00817460819162 & 0.253514 & -0.677215 & 1.57705 & -1.09725 \\ \hline
10 & 2.24678505124783 & 0.249869 & -0.502619 & 0.920542 & -0.403889 \\ \hline
11 & 2.48676496530133 & 0.249348 & -0.474543 & 0.8085 & -0.278266 \\ \hline
12 & 2.72785340315076 & 0.249797 & -0.501405 & 0.921547 & -0.411958 \\ \hline
\multicolumn{1}{|l|}{} & \multicolumn{1}{l|}{} & \multicolumn{1}{l|}{} & \multicolumn{1}{l|}{} & \multicolumn{1}{l|}{} & \multicolumn{1}{l|}{} \\ \hline
\multicolumn{1}{|c|}{$\infty$} & \multicolumn{1}{l|}{} & 0.25 & -0.5 & \multicolumn{1}{l|}{} & \multicolumn{1}{l|}{} \\ \hline
\end{tabular}
\end{center}
\caption{Sixth moment ($n=3$) of the density when $\Gamma=6$.}
\label{mmuN-G6-n3}
\end{table}

%%%%%%%%%%%%%%%%%%%%%%%%%%%%%%%%%%%%%%%%%%%%%
\begin{table}[htbp]
\begin{center}
\begin{tabular}{|r|r|r|r|r|r|}
\hline
\multicolumn{1}{|c|}{$N$} & \multicolumn{1}{c}{$\mmu_N\qquad =$} & \multicolumn{1}{c}{$\quad aN \quad +$} & \multicolumn{1}{c}{$\quad b \quad +$} & \multicolumn{1}{c}{$\quad c/\sqrt{N} \quad +$} & \multicolumn{1}{c|}{$d/N$} \\ \hline
\multicolumn{1}{|l|}{} & \multicolumn{1}{l|}{} & \multicolumn{1}{c|}{$a$} & \multicolumn{1}{c|}{$b$} & \multicolumn{1}{c|}{$c$} & \multicolumn{1}{c|}{$d$} \\ \hline
3 & 0.579848665870171 & \multicolumn{1}{l|}{} & \multicolumn{1}{l|}{} & \multicolumn{1}{l|}{} & \multicolumn{1}{l|}{} \\ \hline
4 & 0.732937257370685 & \multicolumn{1}{l|}{} & \multicolumn{1}{l|}{} & \multicolumn{1}{l|}{} & \multicolumn{1}{l|}{} \\ \hline
5 & 0.897778601877637 & \multicolumn{1}{l|}{} & \multicolumn{1}{l|}{} & \multicolumn{1}{l|}{} & \multicolumn{1}{l|}{} \\ \hline
6 & 1.06814688756782 & 0.178851 & 0.00107475 & -0.227029 & 0.519884 \\ \hline
7 & 1.24220552729734 & 0.18911 & -0.305123 & 0.678947 & -0.231411 \\ \hline
8 & 1.41969193536163 & 0.204817 & -0.8686 & 2.50959 & -1.90014 \\ \hline
9 & 1.60002816650081 & 0.201673 & -0.736901 & 2.04679 & -1.44354 \\ \hline
10 & 1.78261079422029 & 0.198035 & -0.562593 & 1.39137 & -0.751319 \\ \hline
11 & 1.96700538479197 & 0.198245 & -0.573912 & 1.43654 & -0.801964 \\ \hline
12 & 2.15290808747152 & 0.199242 & -0.633669 & 1.68802 & -1.09938 \\ \hline
\multicolumn{1}{|l|}{} & \multicolumn{1}{l|}{} & \multicolumn{1}{l|}{} & \multicolumn{1}{l|}{} & \multicolumn{1}{l|}{} & \multicolumn{1}{l|}{} \\ \hline
\multicolumn{1}{|c|}{$\infty$} & \multicolumn{1}{l|}{} & 0.2 & \multicolumn{1}{l|}{$-2/3\simeq-0.666667$} & \multicolumn{1}{l|}{} & \multicolumn{1}{l|}{} \\ \hline
\end{tabular}
\end{center}
\caption{Eighth moment ($n=4$) of the density when $\Gamma=6$.}
\label{mmuN-G6-n4}
\end{table}

%%%%%%%%%%%%%%%%%%%%%%%%%%%%%%%%%%%%%%%%%%%%%

\begin{table}[htbp]
\begin{center}
\begin{tabular}{|r|r|l|l|l|l|}
\hline
\multicolumn{1}{|c|}{$N$} & \multicolumn{1}{c}{$\mmu_N\qquad =$} & \multicolumn{1}{c}{$\quad aN \quad +$} & \multicolumn{1}{c}{$\quad b \quad +$} & \multicolumn{1}{c}{$\quad c/\sqrt{N} \quad +$} & \multicolumn{1}{c|}{$d/N$} \\ \hline
\multicolumn{1}{|l|}{} & \multicolumn{1}{l|}{} & \multicolumn{1}{c|}{$a$} & \multicolumn{1}{c|}{$b$} & \multicolumn{1}{c|}{$c$} & \multicolumn{1}{c|}{$d$} \\ \hline
3 & 0.691851631655437 &  &  &  &  \\ \hline
4 & 0.992192408359008 &  &  &  &  \\ \hline
5 & 1.30687171332381 &  &  &  &  \\ \hline
6 & 1.62802997072456 & \multicolumn{1}{r|}{0.329038} & \multicolumn{1}{r|}{-0.289875} & \multicolumn{1}{r|}{-0.448497} & \multicolumn{1}{r|}{0.76066} \\ \hline
7 & 1.95072743432764 & \multicolumn{1}{r|}{0.302864} & \multicolumn{1}{r|}{0.491355} & \multicolumn{1}{r|}{-2.7599} & \multicolumn{1}{r|}{2.67751} \\ \hline
\multicolumn{1}{|l|}{} & \multicolumn{1}{l|}{} &  &  &  &  \\ \hline
\multicolumn{1}{|c|}{$\infty$} & \multicolumn{1}{l|}{} & \multicolumn{1}{r|}{$1/3$} & \multicolumn{1}{r|}{-0.5} &  &  \\ \hline
\end{tabular}
\end{center}
\caption{Fourth moment ($n=2$) of the density when $\Gamma=8$.}
\label{mmuN-G8-n2}
\end{table}

%%%%%%%%%%%%%%%%%%%%%%%%%%%%%%%%%%%%%%%%%%%%%

\begin{table}[htbp]
\begin{center}
\begin{tabular}{|r|r|l|l|l|l|}
\hline
\multicolumn{1}{|c|}{$N$} & \multicolumn{1}{c}{$\mmu_N\qquad =$} & \multicolumn{1}{c}{$\quad aN \quad +$} & \multicolumn{1}{c}{$\quad b \quad +$} & \multicolumn{1}{c}{$\quad c/\sqrt{N} \quad +$} & \multicolumn{1}{c|}{$d/N$} \\ \hline
\multicolumn{1}{|l|}{} & \multicolumn{1}{l|}{} & \multicolumn{1}{c|}{$a$} & \multicolumn{1}{c|}{$b$} & \multicolumn{1}{c|}{$c$} & \multicolumn{1}{c|}{$d$} \\ \hline
3 & 0.462978101466508 &  &  &  &  \\ \hline
4 & 0.661530611956197 &  &  &  &  \\ \hline
5 & 0.877122153006864 &  &  &  &  \\ \hline
6 & 1.10026340398747 & \multicolumn{1}{r|}{0.231106} & \multicolumn{1}{r|}{-0.191724} & \multicolumn{1}{r|}{-0.630085} & \multicolumn{1}{r|}{0.97549} \\ \hline
7 & 1.32616206590665 & \multicolumn{1}{r|}{0.215749} & \multicolumn{1}{r|}{0.266649} & \multicolumn{1}{r|}{-1.98632} & \multicolumn{1}{r|}{2.10017} \\ \hline
\multicolumn{1}{|l|}{} & \multicolumn{1}{l|}{} &  &  &  &  \\ \hline
\multicolumn{1}{|c|}{$\infty$} & \multicolumn{1}{l|}{} & \multicolumn{1}{r|}{0.25} & \multicolumn{1}{r|}{-0.75} &  &  \\ \hline
\end{tabular}
\end{center}
\caption{Sixth moment ($n=3$) of the density when $\Gamma=8$.}
\label{mmuN-G8-n3}
\end{table}

%%%%%%%%%%%%%%%%%%%%%%%%%%%%%%%%%%%%%%%%%%%%%

\begin{table}[htbp]
\begin{center}
\begin{tabular}{|r|r|l|l|l|l|}
\hline
\multicolumn{1}{|c|}{$N$} & \multicolumn{1}{c}{$\mmu_N\qquad =$} & \multicolumn{1}{c}{$\quad aN \quad +$} & \multicolumn{1}{c}{$\quad b \quad +$} & \multicolumn{1}{c}{$\quad c/\sqrt{N} \quad +$} & \multicolumn{1}{c|}{$d/N$} \\ \hline
\multicolumn{1}{|l|}{} & \multicolumn{1}{l|}{} & \multicolumn{1}{c|}{$a$} & \multicolumn{1}{c|}{$b$} & \multicolumn{1}{c|}{$c$} & \multicolumn{1}{c|}{$d$} \\ \hline
3 & 0.359389450389747 &  &  &  &  \\ \hline
4 & 0.499919485631197 &  &  &  &  \\ \hline
5 & 0.656534473821437 &  &  &  &  \\ \hline
6 & 0.819712345574349 & \multicolumn{1}{r|}{0.165419} & \multicolumn{1}{r|}{0.0211561} & \multicolumn{1}{r|}{-0.961295} & \multicolumn{1}{r|}{1.19094} \\ \hline
7 & 0.985867782818619 & \multicolumn{1}{r|}{0.16514} & \multicolumn{1}{r|}{0.0294668} & \multicolumn{1}{r|}{-0.985884} & \multicolumn{1}{r|}{1.21133} \\ \hline
\multicolumn{1}{|l|}{} & \multicolumn{1}{l|}{} &  &  &  &  \\ \hline
\multicolumn{1}{|c|}{$\infty$} & \multicolumn{1}{l|}{} & \multicolumn{1}{r|}{0.2} & \multicolumn{1}{r|}{-1} &  &  \\ \hline
\end{tabular}
\end{center}
\caption{Eighth moment ($n=4$) of the density when $\Gamma=8$.}
\label{mmuN-G8-n4}
\end{table}
%%%%%%%%%%%%%%%%%%%%%%%%%%%%%%%%%%%

\break

\section{Conclusion}
\label{S6}

In this work we explored two applications to the 2dOCP of the expansion
of powers of the Vandermonde determinant based on the formalism
presented in section \ref{S2}. The first is the study of the moments of
the pair correlation function on the sphere. We showed that the second
moment satisfies an exact relation for finite number of particles $N$,
and we explored numerically the behavior of higher order moments. Also
an approximation to these moments using a formalism based on the
direct correlation function was proposed.

The second result is on the evaluation of the distribution function of
the linear statistic $\sum_{l=1}^N |\vec{r}_l|^{2n}$ in the soft disk
geometry. The exact distribution tends to a Gaussian when
$N\to\infty$. We were able to compute the $\OO(1)$ correction to mean
of this linear statistic for any $n$, and deduce it for a general
smooth radial linear statistic. The result was checked explicitly at
$\Gamma=2$, for values of $\Gamma$ close to 2, and numerically for
$\Gamma=4$ and 6.

\section*{Acknowledgements}

We thank N.~Regnault for providing us with the code to compute
numerically the coefficients $\{c_{\mu}^{(N)}\}$ based
on~(\ref{eq:cmu-boson}) and~(\ref{eq:cmu-fermion}).

GT acknowledges partial financial support from Facultad de Ciencias
de la Universidad de los Andes, and ECOS Nord/COLCIENCIAS-MEN-ICETEX.
PJF was supported by the Australian Research Council.

\section*{Appendix}
\setcounter{equation}{0}
\renewcommand{\theequation}{A.\arabic{equation}}

In this appendix we present some properties of the functions
\begin{equation}
  \label{eq:I}
  \I(k_1,k_2)=\iint_{0\leq t_2 < t_1} e^{-t_1-t_2} t_1^{k_1} t_2^{k_2}
  \,dt_1dt_2
  \,,
\end{equation}
and
\begin{equation}
  \J(k_1,k_2)=\int_{0}^{\infty}\int_{0}^{\infty}
  e^{-t_1-t_2} t_1^{k_1} t_2^{k_2} \log(\max(t_1 ,t_2)) \,dt_1dt_2
  \,,
\end{equation}
that appear in the expansion of the density around $\Gamma=2$. First,
it should be noticed that
\begin{equation}
  \label{eq:Isymmetry}
  \I(k_1,k_2)+  \I(k_2,k_1)= k_1! k_2!\ .
\end{equation}
Doing an integration by parts, one obtains the recurrence relation
\begin{equation}
  \I(k_1+1,k_2)-(k_1+1)\I(k_1,k_2)=2^{-k_1-k_2-2}(k_1+k_2+1)!
  \,,
\end{equation}
and reiterating
\begin{equation}
  \label{eq:Irec}
   \I(k_1+n,k_2)-\frac{(k_1+n)!}{k_1!}\,\I(k_1,k_2)=
   \sum_{\ell=0}^{n-1}2^{-k_1-k_2-2-\ell}\frac{(k_1+k_2+1+\ell)!(k_1+n)!}{(k_1+\ell)!}
   \,.
\end{equation}
Similarly, for $\J$ we have
\begin{equation}
  \label{eq:Jrec}
  \J(k_1+n,k_2)-\frac{(k_1+n)!}{k_1!}\,\J(k_1,k_2)=
   \sum_{\ell=0}^{n-1} \I(k_1+\ell,k_2)\frac{(k_1+n)!}{(k_1+\ell+1)!}
   \,.  
\end{equation}
From (\ref{eq:Irec}), one can obtain an alternative
expression for $\I$ as a sum
\begin{equation}
  \I(k_1,k_2)=\sum_{\ell=0}^{k_1} 2^{-k_2-\ell-1} \frac{(k_2+\ell)!k_1!}{\ell!}\,.
\end{equation}

The asymptotic expansion of $\I(k_1,k_2)$ for large arguments, $k_1$
and $k_2$ of order $N\to\infty$, can be obtained by the steepest
descent method. The maximum of the integrand in~(\ref{eq:I}) is for
$t_1=k_1$ and $t_2=k_2$. Therefore, the behavior of $\I(k_1,k_2)$ will
depend on whether this maximum is in the domain of integration $0\leq
t_2<t_1$ or not, i.e., if $k_2<k_1$ or not. If $1\ll k_2<k_1$, and
$|k_1-k_2|/\sqrt{N}\gg 1$, then the maximum of the integrand
in~(\ref{eq:I}) is deep inside the domain of integration, and a simple
application of the steepest descent shows that $\I(k_1,k_2)\sim k_1!
k_2!\,$. However for the calculations of section~\ref{SsubGamma2}, the
behavior of $\I(k_1,k_2)$ when $k_2>k_1$ is needed. In this situation,
the maximum of the integrand is outside the domain of
integration. Nevertheless, $\I(k_1,k_2)$ will give a significant
contribution when the maximum of the integrand is ``close'' to the
border, more precisely when $|k_2-k_1|$ is of order $O(\sqrt{N})$. The
dominant contribution to the integral~(\ref{eq:I}) will be given by
the region consisting of a strip attached and parallel to the line
$t_1=t_2$ of width of order $\sqrt{N}$. To be more specific, let us do
the change of integration variables $v_{+}=t_1+t_2$ and
$v_{-}=t_1-t_2$ in (\ref{eq:I}) and write
\begin{equation}
  \I(k_1,k_2)=\iint_{\cal D} dv_{+}dv_{-} e^{-g(v_+,v_{-})} \frac{dv_{+}dv_{-}}{2^{k_1+k_2+1}}
\,,
\end{equation}
with
\begin{equation}
  g(v_{+},v_{-})=v_{+}-k_1\log(v_{+}+v_{-})-k_2\log(v_{+}-v_{-})
\,,
\end{equation}
where ${\cal D}=\{(t_1,t_2), 0\leq t_2 < t_1\}$. $g$ has its maximum
for $v_{+}=v_{+}^*=k_1+k_2$ and $v_{-}=v_{-}^*=k_1-k_2$ (i.e. $t_1=k_1$
and $t_2=k_2$). Expanding $g$ to the second order around its maximum
we obtain
\begin{equation}
  \label{eq:Igaussian}
  \I(k_1,k_2)\sim \frac{e^{-k_1-k_2} k_1^{k_1} k_2^{k_2}}{2}
  \iint_{\cal D}
e^{ -\frac{k_1+k_2}{8k_1k_2}\left[(v_{+}-v_{+}^{*})^2
+(v_{-}-v_{-}^{*})^2\right]-\frac{k_2-k_1}{4k_1k_2}(v_{+}-v_{+}^{*})(v_{-}-v_{-}^{*})
}\,dv_{+}dv_{-}
\,.
\end{equation}
Although the domain of integration ${\cal D}$ cannot be simply
expressed in terms of the variables $v_{+}$ and $v_{-}$, due to fast
Gaussian decay of the integrand in~(\ref{eq:Igaussian}), the dominant
contribution is indeed obtained by the strip along the line $t_1=t_2$
mentioned above. With this in mind, it is clear that one can extend
the domain of integration for $v_{+}$ to $\left]-\infty,
+\infty\right[$ and for $v_{-}$ to $\left[0,+\infty\right[$, up to
exponentially small corrections. Then, performing first the
integral over $v_{+}$, we obtain
\begin{equation}
\I(k_1,k_2)\sim  e^{-k_1-k_2} k_1^{k_1} k_2^{k_2} \sqrt{\frac{2\pi
    k_1k_2}{k_1+k_2}}
\int_{0}^{+\infty} e^{-\frac{(v_{-}-v_{-}^{*})^2}{2(k_1+k_2)}}
  \,dv_{-}\,.
\end{equation}
Performing now the integral over $v_{-}$ gives
\begin{equation}
  \label{eq:Isim}
  \I(k_1,k_2) \sim e^{-k_1-k_2} k_1^{k_1} k_2^{k_2} \pi \sqrt{k_1k_2}\,
  \erfc\left(\frac{k_2-k_1}{\sqrt{2(k_1+k_2)}}\right)
  \,,
\end{equation}
where $\erfc(x)=1-(2/\sqrt{\pi})\int_{0}^{x} e^{-t^2}\,dt$, is the
complementary error function. Recalling Stirling's formula for the
factorial, (\ref{eq:Isim}) can be written as
\begin{equation}
  \label{eq:Iasympt}
   \frac{\I(k_1,k_2)}{ k_1! k_2! }  \sim 
   \frac{1}{2}\,\erfc\left(\frac{k_2-k_1}{\sqrt{2(k_1+k_2)}}\right)
   \,.
\end{equation}

%\bibliographystyle{amsplain}
%\bibliography{book1}

\providecommand{\bysame}{\leavevmode\hbox to3em{\hrulefill}\thinspace}
\providecommand{\MR}{\relax\ifhmode\unskip\space\fi MR }
\providecommand{\MRhref}[2]{%
  \href{http://www.ams.org/mathscinet-getitem?mr=#1}{#2}
}
\providecommand{\href}[2]{#2}

\end{document}